\documentclass[preprint,showpacs,preprintnumbers,amsmath,amssymb,nofootinbib]{revtex4}
\usepackage{graphicx}

\usepackage{dcolumn}
\usepackage{bm}

\begin{document}
\title{The White Dwarf --- White Dwarf galactic background in the LISA
  data.}

\author{Jeffrey A. Edlund}
\email{Jeffrey.A.Edlund@jpl.nasa.gov}
\affiliation{Jet Propulsion Laboratory, California Institute of
             Technology, Pasadena, CA 91109}

\author{Massimo Tinto}
\email{Massimo.Tinto@jpl.nasa.gov}
\affiliation{Jet Propulsion Laboratory, California Institute of
             Technology, Pasadena, CA 91109}

\author{Andrzej Kr\'olak}
\email{krolan@aei.mpg.de} \altaffiliation[also at: ]{Institute of
Mathematics, Polish Academy of Sciences, Warsaw, Poland}
\affiliation{Max-Planck-Institute for Gravitational Physics,
Albert Einstein Institute, D-14476 Golm, Germany}

\author{Gijs Nelemans} 
\email{nelemans@astro.ru.nl}
\affiliation{Department of Astrophysics, IMAPP, Radboud University
  Nijmegen, The Netherlands}

\date{\today}
\begin{abstract}
  LISA (Laser Interferometer Space Antenna) is a proposed space
  mission, which will use coherent laser beams exchanged between three
  remote spacecraft to detect and study low-frequency cosmic
  gravitational radiation. In the low-part of its frequency band, the
  LISA strain sensitivity will be dominated by the incoherent
  superposition of hundreds of millions of gravitational wave signals
  radiated by inspiraling white-dwarf binaries present in our own
  galaxy. In order to estimate the magnitude of the LISA response to
  this background, we have simulated a synthesized population that
  recently appeared in the literature. Our approach relies on entirely
  analytic expressions of the LISA Time-Delay Interferometric
  responses to the gravitational radiation emitted by such systems,
  which allows us to implement a computationally efficient and
  accurate simulation of the background in the LISA data. We find the
  amplitude of the galactic white-dwarf binary background in the LISA
  data to be modulated in time, reaching a minimum equal to about
  twice that of the LISA noise for a period of about two months around
  the time when the Sun-LISA direction is roughly oriented towards the
  Autumn equinox.  This suggests that, during this time period, LISA
  could search for other gravitational wave signals incoming from
  directions that are away from the galactic plane.  Since the
  galactic white-dwarfs background will be observed by LISA not as a
  stationary but rather as a cyclostationary random process with a
  period of one year, we summarize the theory of cyclostationary
  random processes, present the corresponding generalized spectral
  method needed to characterize such process, and make a comparison
  between our analytic results and those obtained by applying our
  method to the simulated data. We find that, by measuring the
  generalized spectral components of the white-dwarf background, LISA
  will be able to infer properties of the distribution of the
  white-dwarfs binary systems present in our Galaxy.
\end{abstract}

\maketitle

\section{Introduction}
\label{intro}

The Laser Interferometric Space Antenna (LISA) is a space mission
jointly proposed to the National Aeronautics and Space Administration
(NASA) and the European Space Agency (ESA). Its aim is to detect and
study gravitational waves (GW) in the millihertz frequency band. It
will use coherent laser beams exchanged between three identical
spacecraft forming a giant (almost) equilateral triangle of side $5
\times 10^6$ kilometers. By monitoring the relative phase changes of
the light beams exchanged between the spacecraft, it will extract the
information about the gravitational waves it will observe at
unprecedented sensitivities  \cite{PPA98}.

The astrophysical sources that LISA is expected to observe within its
operational frequency band ($10^{-4} - 1$ Hz) include extra-galactic
super-massive black-hole coalescing binaries, stochastic gravitational
wave background from the early universe, and galactic and
extra-galactic coalescing binary systems containing white dwarfs and
neutron stars.

Recent surveys have uniquely identified twenty binary systems emitting
gravitational radiation within the LISA band, while population studies
have concluded that the large number of binaries present in our own
galaxy should produce a stochastic background that will lie
significantly above the LISA instrumental noise in the low-part of its
frequency band.  It has been shown in the literature (see \cite{NYP01}
for a recent study and \cite{HBW90,EIS87} for earlier investigations)
that these sources will be dominated by detached white-dwarf ---
white-dwarf (WD-WD) binaries, with $1.1 \ \times 10^8$ of such systems
in our Galaxy. The detached WD-WD binaries evolve by
gravitational-radiation reaction and the number of such sources
rapidly decreases with increasing orbital frequency. Although it is
expected that, above a certain frequency cut-off ($1 - 2 \ {\rm
  mHz}$), we will be able to resolve individual signals and remove
them from the LISA data, it is still not clear how to further improve
the LISA sensitivity to other gravitational wave signals in the region
of the frequency band below the WD-WD background frequency cut-off.
Although two promising data analysis procedures have been proposed
\cite{CL03,KT03} for attempting to subtract the galactic background,
considerable work still needs to be done to verify their
effectiveness. In this context, simulating the LISA response to the
WD-WD background will be particularly useful for verifying present and
future data analysis ``cleaning'' algorithms.  A realistic simulation
will also quantify the effects of the LISA motion around the Sun on
the overall amplitude and phase of the GW signal generated by the
background in the LISA data.  The directional properties of the LISA
response and its time dependence introduced by the motion of LISA
around the Sun, together with the non-isotropic and non-homogeneous
distribution of the WD-WD binary systems within the galactic disk as
seen by LISA, imply that the magnitude of the background observed by
LISA will not be a stationary random process. As a consequence of the
one-year periodicity of the LISA motion around the Sun, there exist
relatively long ($\approx 2$ months) stretches of data during which
the magnitude of the LISA response to the background will reach an
absolute minimum \cite{Seto04}. Our simulation shows this minimum to
be less than a factor of two larger than the level identified by the
LISA secondary noises, suggesting the possibility of performing
searches for gravitational radiation from other sources located in
regions of the sky that are away from the galactic plane.  The LISA
sensitivity to such signal in fact will be less limited by the WD-WD
background during these periods of observation.

This paper is organized as follows. In Section \ref{LISA_RES} we
provide the analytic expression of one of the LISA Time-Delay
Interferometric (TDI) responses to a signal radiated by a binary
system. Although all the TDI responses to binary signals were first
given in their closed analytic form in \cite{KTV04}, in what follows
we will focus our attention only on the unequal-arm Michelson
combination, $X$. In Section \ref{WD_POP} we give a summary of how the
WD-WD binary population was obtained, and a description of our
numerical simulation of the $X$ response to it.  In Section \ref{Simu}
we describe the numerical implementation of our simulation of the LISA
$X$ response to the WD-WD background, and summarize our results.  In
particular, in agreement with the results by Seto \cite{Seto04}, we
find the amplitude of the galactic WD-WD background in the LISA
$X$-combination to be modulated in time, reaching a minimum when the
Sun-LISA direction is roughly oriented towards the Autumn equinox.
Furthermore, we show the amplitude of the background at its minimum to
be a factor less than two larger than the level identified by the LISA
noise for a time period of about two months, suggesting that LISA
could search (during this time period) for other gravitational wave
signals incoming from regions of the sky that are away from the
galactic plane.

The time-dependence and periodicity of the magnitude of the WD-WD
galactic background in the LISA data implies that it is not a
stationary but rather a cyclostationary random process of period one
year.  After providing a brief summary of the theory of
cyclostationary random processes relevant to the LISA detection of the
WD-WD galactic background, we apply it to three years worth of
simulated LISA $X$ data. We find that, by measuring the generalized
spectral components of such cyclostationary random process, LISA will
be able to infer key-properties of the distribution of the WD-WD
binary systems present in our own Galaxy.

\section{The LISA response to signals from binary systems}
\label{LISA_RES}

\begin{figure}
\centering
\includegraphics[width=2.5in, angle=-90]{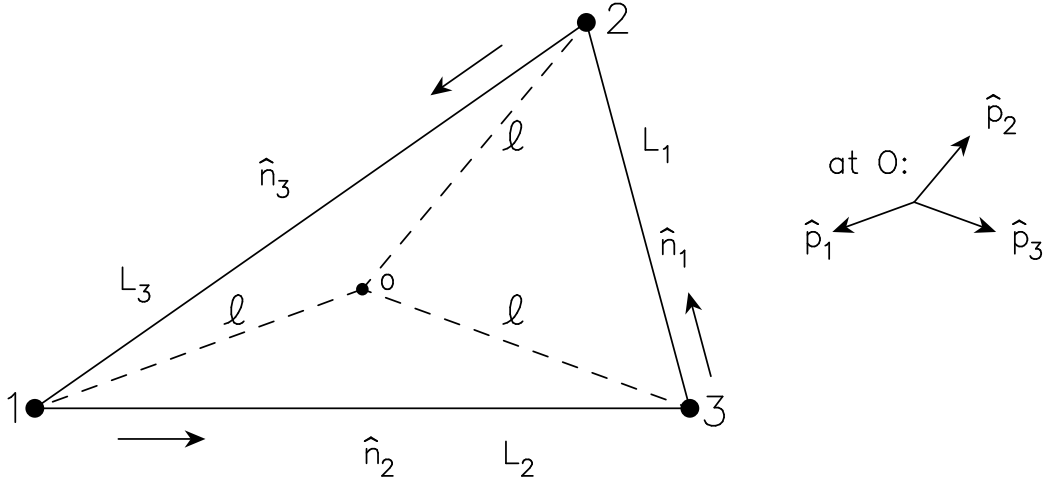}
\caption{Schematic LISA configuration. Each spacecraft is 
  equidistant from point $o$, with unit vectors $\hat p_i$ indicating
  directions to the three spacecraft. Unit vectors $\hat n_i$ point
  between spacecraft pairs with the indicated orientation. \label{Fig1}}
\end{figure}

The overall LISA geometry is shown in Figure (\ref{Fig1}).
There are six beams exchanged between the LISA spacecraft, 
with the six Doppler measurements $y_{ij}$ ($i,j = 1, 2, 3$) recorded
when each received beam is mixed with the laser light of the
receiving optical bench.  The frequency fluctuations from the six
lasers, which enter in each of the six Doppler measurements, need to
be suppressed to a level smaller than that identified by the secondary
 (proof mass and optical path) noises  \cite{TEA02} in order to detect
and study gravitational radiation at the predicted amplitudes. 

\begin{figure}
\centering
\includegraphics[width=3.0 in, angle=0.0]{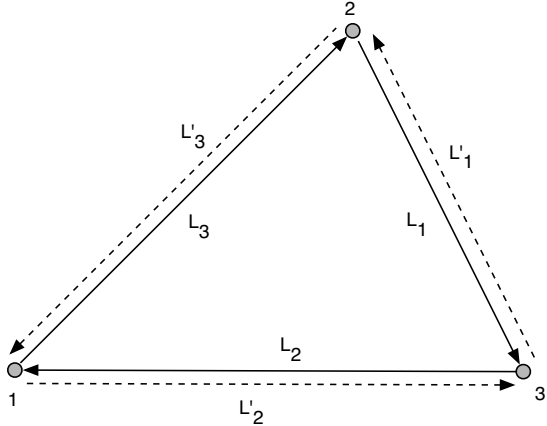}
\caption{Schematic diagram of LISA configurations involving
  six laser beams. Optical path delays taken in the counter-clockwise
  sense are denoted with a prime, while unprimed delays are in the
  clockwise sense. 
\label{Fig2}}
\end{figure}

Since the LISA triangular array has systematic motions, the two
one-way light times between any spacecraft pair are not the same
 \cite{S04}.  Delay times for light travel between
the spacecraft must be accounted for depending on the sense of light
propagation along each link when combining these data as a consequence
of the rotation of the array.  Following  \cite{TEA04}, the arms are
labeled with single numbers given by the opposite spacecraft; e.g.,
arm $2$ (or $2^{'}$) is opposite spacecraft $2$, where primed delays are
used to distinguish light-times taken in the counter-clockwise sense
and unprimed delays for the clockwise light times (see Figure (\ref{Fig2})).
Also the following labeling convention of the Doppler data will be
used.  Explicitly: $y_{23}$ is the one-way Doppler shift measured at
spacecraft $3$, coming from spacecraft $2$, along arm $1$.  Similarly,
$y_{32}$ is the Doppler shift measured on arrival at spacecraft $2$
along arm $1'$ of a signal transmitted from spacecraft $3$.  Due to the
relative motion between spacecraft, $L_1 \neq L_1^{'}$ in general. As
in  \cite{ETA00,TEA02}, we denote six further data streams,
$z_{ij}$ ($i,j = 1, 2, 3$), as the intra-spacecraft metrology data
used to monitor the motion of the two optical benches and the relative
phase fluctuations of the two lasers on each of the three spacecraft.
The frequency fluctuations introduced by the lasers, by the optical
benches, by the proof masses, by the fiber optics, and by the
measurements themselves at the photo-detectors (i.e.\ the shot-noise
fluctuations) enter the Doppler observables $y_{ij}$, $z_{ij}$ with
specific time signatures; see Refs.  \cite{ETA00,TEA02} for a detailed
discussion.  The contribution $y^\mathrm{GW}_{ij}$ due to GW signals
was derived in Ref.  \cite{AET99} in the case of a stationary array,
and further extended to the realistic configuration \cite{KTV04} of
the LISA array orbiting around the Sun.

Let us consider for instance the ``second generation'' unequal-arm
Michelson TDI observables, ($X_1, X_2, X_3$). Their expressions, in
terms of the Doppler measurements $y_{ij}$, $z_{ij}$, are as follows \cite{STEA03}
\begin{eqnarray}
X_1 & = & [{(y_{31} + y_{13;2}) + (y_{21} + y_{12;3'})}_{;2'2}
+ {(y_{21} + y_{12;3'})}_{;33'2'2}
+ {(y_{31} + y_{13;2})}_{;33'33'2'2}]
\nonumber
\\
& & -
[(y_{21} + y_{12;3'})
+ {(y_{31} + y_{13;2})}_{;33'}
+ {(y_{31} + y_{13;2})}_{2'233'} +
{(y_{21} + y_{12;3'})}_{;2'22'233'}]
\nonumber
\\
& & +  \frac{1}{2} \ [(z_{21} - z_{31}) - {(z_{21} -
z_{31})}_{;33'} - {(z_{21} - z_{31})}_{;2'2}
+ {(z_{21} - z_{31})}_{;33'33'2'2}
\nonumber
\\
& & + {(z_{21} - z_{31})}_{;2'22'233'}
- {(z_{21} - z_{31})}_{;2'233'33'2'2}] \ ,
\label{eq:X1}
\end{eqnarray}
with $X_2$, $X_3$ following from Eq. (\ref{eq:X1}) by permutations of
the spacecraft indices. The semicolon notation shown in equation
(\ref{eq:X1}) emphasizes that the operation of sequentially applying
two or more delays to a given measurement is non-commutative as
consequence of the time dependence of the light-times $L_i$ and
$L_i^{'}$ ($i = 1, 2, 3$), and a specific order has to be adopted to
adequately suppress the laser noises \cite{TEA04,CH03,STEA03}.
Specifically: $y_{ij;kl} \equiv y_{ij} (t - L_l (t) - L_k (t - L_l))
\ne y_{ij;lk}$ (units in which the speed of light $c=1$).

The expressions of the gravitational wave signal and the secondary
noise sources entering into $X_1$ will in general be different from
those entering into $X$, the corresponding ``first generation''
unequal-arm Michelson observable derived under the assumption of a
stationary LISA array \cite{AET99,ETA00}.  However, the magnitude
of the corrections introduced by the motion of the array are
proportional to the product between the time derivative of the GW
amplitude and the difference between the actual light travel times and
those valid for a stationary array.  At $1$ Hz, for instance, the
larger correction to the signal (due to the difference between the
co-rotating and counter-rotating light travel times) is two orders of
magnitude smaller than the main signal. Since the amplitude of this
correction scales linearly with the Fourier frequency, we can
completely disregard this effect (and the weaker effect due to the
time dependence of the light travel times) over the entire LISA band
 \cite{TEA04}.  Furthermore, since along the LISA orbit the three
armlengths will differ at most by $\sim$ 1\%--2\%, the degradation in
signal-to-noise ratio introduced by adopting signal templates that
neglect the inequality of the armlengths will be of only a few
percent.  For these reasons, in what follows we will focus on the
expressions of the GW responses of various second-generation Time-Delay
Interferometry (TDI) observables by disregarding the differences in
the delay times experienced by light propagating clockwise and
counterclockwise, and by assuming the three LISA armlengths to be
constant and equal to $L = 5 \times 10^6 \, \mbox{km} \simeq 16.67 \,
\mbox{s}$ \cite{PPA98}.  These approximations, together with the
treatment of the moving-LISA GW response discussed in \cite{KTV04} are
essentially equivalent to the \emph{rigid adiabatic approximation\/} of
Ref. \cite{RCP04}, and to the formalism of Ref. \cite{Seto04}.

These considerations imply that the second generation TDI expressions
for the gravitational wave signal and the secondary noises can be
expressed in terms of the corresponding first generation TDIs.  For
instance, the gravitational wave signal entering into the second
generation unequal-arm Michelson combination, $X^{\rm GW}_1$, can be
written in terms of the gravitational wave response of the
corresponding first generation unequal-arm Michelson combination,
$X^{\rm GW} (t)$, in the following manner \cite{TL04}
\begin{equation}
X^{\rm GW}_1 (t) = X^{\rm GW} (t) - X^{\rm GW} (t - 4L)
\label{X1fromX}
\end{equation}
Equation (\ref{X1fromX}) implies that any data analysis procedure and
algorithm that will be implemented for the second generation TDI
combinations can actually be derived by considering the corresponding
first generation TDI expressions. For this reason, from now on we
will focus our attention on the gravitational wave responses of the
first generation combinations.

The gravitational wave response $X^{\rm GW} (t)$ of the unequal-arm
Michelson TDI combination to a signal from a binary system has been
derived in \cite{KTV04}, and it can be written in the following form
\begin{equation}
X^\mathrm{GW} (t) = \Re \left[A(x, t) \ e^{-i \phi (t)}\right] \ ,
\label{X}
\end{equation}
where $x = \omega_s L$ ($\omega_s$ being the angular frequency of the
GW signal in the source reference frame), and the expressions for the
complex amplitude $A (x, t)$ and the real phase $\phi (t)$
are
\begin{eqnarray}
A(x, t) & = & 2 \, x \, \sin(x) \left\{ \left[ sinc[(1+c_2(t))\frac{x}{2}] \ e^{i x(\frac{3}{2} +
  d_2 (t))} 
+ sinc[(1-c_2(t))\frac{x}{2}] \ e^{i x(\frac{5}{2} +  d_2 (t))}
  \right] \right. \ {\cal B}_2 (t) 
\nonumber
\\
& - & \left. \left[ sinc[(1-c_3(t))\frac{x}{2}] \ e^{i x(\frac{3}{2} +
  d_3 (t))} 
+ sinc[(1+c_3(t))\frac{x}{2}] \ e^{i x(\frac{5}{2} +  d_3 (t))}
  \right] \ {\cal B}_3 (t) \right\} \ ,
\label{A}
\end{eqnarray}
\begin{equation}
\phi(t) = \omega_s t + \omega_s \ R \ \cos \beta \ \cos(\omega_s t +
\eta_0 - \lambda) \ .
\label{phi}
\end{equation}
In equation (\ref{phi}) $R$ is the distance of the guiding center of
the LISA array, $o$, from the Solar System Barycenter, ($\beta,
\lambda$) are the ecliptic latitude and longitude respectively of the
source location in the sky, $\Omega = 2 \pi/{\rm year}$, and
$\eta_0$ defines the position of the LISA guiding center in the
ecliptic plane at time $t = 0$. Note that the functions $c_k (t)$,
$d_k (t)$, and ${\cal B}_k (t)$ ($k = 2, 3$) do not depend on $x$. The
analytic expressions for $c_k (t)$, and $d_k (t)$ are the same as
those given in equations (46,47) of reference \cite{KTV04}, while the
functions ${\cal B}_k (t)$ ($k = 2, 3$) are equal to
\begin{equation}
{\cal B}_k (t) = (a^{(1)} + i \ a^{(3)}) \ u_k (t) + (a^{(2)} + i \ a^{(4)}) \ v_k
(t) \ .
\label{B}
\end{equation}
The coefficients ($a^{(1)}, a^{(2)}, a^{(3)}, a^{(4)}$) depend only on
the two independent amplitudes of the gravitational wave signal,
($h_+$, $h_\times$), the polarization angle, $\psi$, and an arbitrary
phase, $\phi_0$, that the signal has at time $t = 0$. Their analytic
expressions are given in equations (41--44) of reference \cite{KTV04},
while the functions $u_k (t)$, and $v_k (t)$ ($k=2, 3$) are given in
equations (27,28) in the same reference.

Since most of the gravitational wave energy radiated by the galactic
WD-WD binaries will be present in the lower part of the LISA
sensitivity frequency band, say between $10^{-4} - 10^{-3}$ Hz, it is
useful to provide an expression for the Taylor expansion of the $X$
response in the long-wavelength limit (LWL), i.e. when the wavelength
of the gravitational wave signal is much larger than the LISA
armlength ($x << 1$). As it will be shown in the following sections,
the LWL expression will allow us to analytically describe the general
features of the white dwarfs background in the $X$-combination, and
derive computationally efficient algorithms for numerically simulating
the WD-WD background in the LISA data.

The nth-order truncation, $X^{\rm GW}_{(n)} (t)$, of the Taylor
expansion of $X^{\rm GW}(t)$ in power series of $x$ can be written in
the following form
\begin{equation}
X^{\rm GW}_{(n)} (t) = Re \sum_{k=0}^n A^{(k)} (t) \ x^{k+2} \ e^{-i \phi (t)}
\ ,
\label{Xn}
\end{equation}
where the first three functions of time $A^{(k)} (t), \ \ k \le 2$
are equal to
\begin{eqnarray}
A^{(0)}  & = &  4 \ [{\cal B}_2 - {\cal B}_3 ] \ ,
\nonumber
\\
A^{(1)}  & = & 4 i \ [(d_2+2) \ {\cal B}_2  - (d_3+2) {\cal B}_3 ] \ ,
\nonumber
\\
A^{(2)}  & = & [2{d_3}^2  + 8 d_3 + \frac{28}{3} + \frac{1}{6}
{c_3}^2] \ {\cal B}_3 - [2{d_2}^2  + 8 d_2 + \frac{28}{3} + \frac{1}{6}
{c_2}^2] \ {\cal B}_2 \ .
\label{An}
\end{eqnarray}
Note that the form we adopted for $X^{\rm GW} (t)$ (equation \ref{X})
makes the derivation of the functions $A^{(k)} (t)$ particularly easy
since the dependence on $x$ in $A (x,t)$ is now limited only to the
coefficients in front of the two functions ${\cal B}_2 (t)$ and ${\cal B}_3 (t)$
(see equation (\ref{A})). 

\begin{figure}
\includegraphics[width=5in]{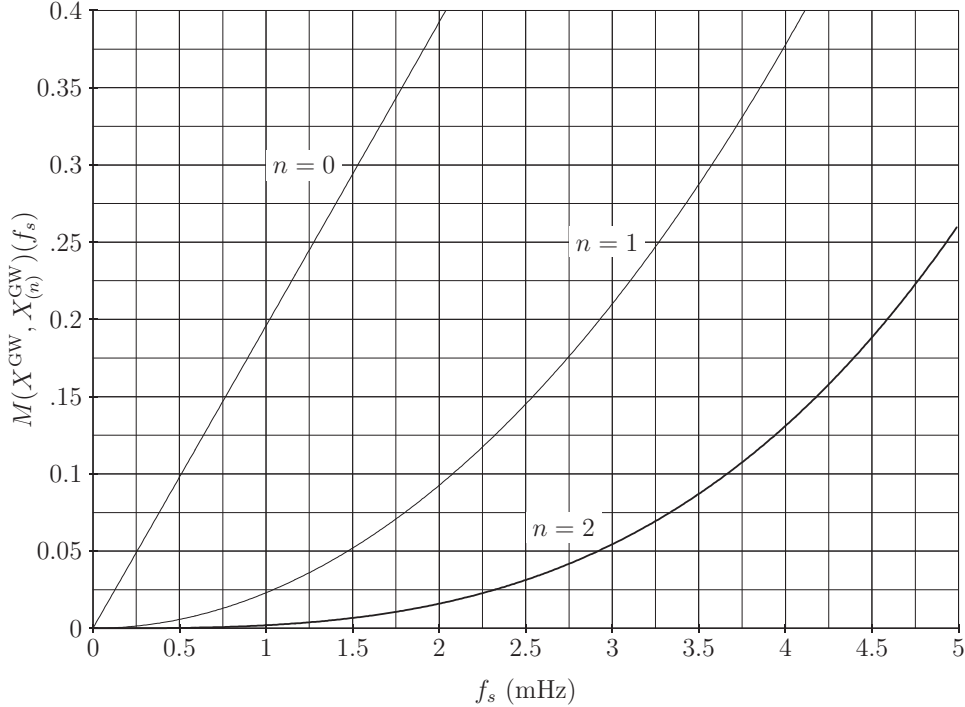}
\caption{Plots of the percentage root-mean-squared errors, $
  M(X^{\rm GW}, X^{\rm GW}_{(n)})$, associated with the
  long-wavelength expansion index $n$, as functions of the
  gravitational wave frequency, $f_s$. The source location has been
  assumed to be in the center of our galaxy.}
\label{error}
\end{figure}

Although it is generally believed that the lowest order
long-wavelength expansion of the $X$ combination, $X^{\rm GW}_{(0)}$,
is sufficiently accurate in representing a gravitational wave signal
in the low-part of the LISA frequency band, there has not been in the
literature any quantitative analysis of the error introduced by
relying on such a zero-order approximation.  Since any TDI combination
will contain a linear superposition of tens of millions of signals, it
is crucial to estimate such an error as a function of the order of the
approximation, $n$. In order to determine how many terms we need to
use for a given signal angular frequency, $\omega_s$, we will rely on
the following `` matching function''
\begin{equation}
M(X^{\rm GW}, X^{\rm GW}_{(n)}) \equiv  \sqrt{\frac{\int_{0}^{T} {[X^{\rm GW} (t) - X^{\rm
    GW}_{(n)} (t)]}^2 dt}{\int_{0}^{T} {[X^{\rm GW} (t)]}^2 dt}} \ .
\label{M}
\end{equation}
Equation (\ref{M}) estimates the percent root-mean-squared error
implied by using the $n^{\rm th}$ order LWL approximation. In Figure
(\ref{error}) we plot $M$ as a function of the signal frequency,
$f_s$ ($=\omega_s/2\pi$), for $n=0, 1, 2$. At $5 \times 10^{-4}$ Hz, for
instance, the zero-order LWL approximation ($n = 0$) of the $X$
combination shows an r.m.s.\ deviation from the exact response equal
to about $10$ percent. As expected, this inaccuracy increases for
signals of higher frequencies, becoming equal to $40$ percent at $2
\times 10^{-3}$ Hz. With $n=1$ the accuracy improves showing that the
$X^{\rm GW}_{(1)}$ response deviates from the exact one with an r.m.s.
error smaller than $10$ percent in the frequency band ($10^{-4} - 2
\times 10^{-3}$) Hz. In our simulation we have actually implemented
the $n=2$ LWL expansion because it was possible and easy to do.

\section{White Dwarf binary population distribution}
\label{WD_POP}

The gravitational wave signal radiated by a WD-WD binary system
depends on eight parameters, ($\phi_o, \iota, \psi, D, \beta, \lambda,
{\cal M}_c, \omega_s$), which are the constant phase of the signal
 ($\phi_o$) at the starting time of the observation, the inclination
angle ($\iota$) of the angular momentum of the binary system relative
to the line of sight, the polarization angle ($\psi$) describing the
orientation of the wave polarization axes, the distance ($D$) to the
binary, the angles ($\lambda, \beta$) describing the location of the
source in the sky relative to the ecliptic plane, the chirp mass
 (${\cal M}_c$), and the angular frequency ($\omega_s$) in the source
reference frame respectively.  Since it can safely be assumed that the
chirp mass ${\cal M}_c$ and the angular frequency $\omega_s$ are
independent of the source location \cite{NYP01} and of the remaining
angular parameters $\phi_o, \iota, \psi$, and because there are no
physical arguments for preferred values of the constant phase $\phi_o$
and the orientation of the binary given by the angles $\iota$ and
$\psi$, it follows that the joint probability distribution, $P(\phi_o,
\iota, \psi, D, \beta, \lambda, {\cal M}_c, \omega_s)$, can be
rewritten in the following form
\begin{equation}
P(\phi_o, \iota, \psi, D, \beta, \lambda, {\cal M}_c, \omega_s) = 
P_1 (\phi_o) P_2 (\iota) P_3 (\psi) P_4 (D, \beta, \lambda) 
P_5({\cal M}_c, \omega_s) \ . 
\end{equation}
In the implementation of our simulation we have assumed the angles
$\phi_o$ and $\psi$ to be uniformly distributed in the interval $[0, 2
\pi)$, and $\cos\iota$ uniformly distributed in the interval $[-1,1]$.
We further assumed the binary systems to be randomly distributed in
the Galactic disc according to the following axially symmetric
distribution ${\cal P}_4 (R, z)$ (see \cite{NYP01} Eq. (5))
\begin{equation}
{\cal P}_4 (R,z) = \frac{e^{-R/H} \ sech^2(z/z_o)}{4 \pi z_o H^2}  \ ,
\label{eq:Galdis}
\end{equation}
where $(R,z)$ are cylindrical coordinates with origin at the galactic
center, $H = 2.5$ kpc, and $z_o = 200$ pc, and it is proportional to
$P_4 (D, \lambda, \beta)$ through the Jacobian of the coordinate
transformation.  Note that the position of the Sun in this coordinate
system is given by $R_{\odot} = 8.5 \ {\rm kpc}$ and $z_{\odot} = -30
\ {\rm pc}$. We then generate the positions of the sources from the
distribution given by Eq.  (\ref{eq:Galdis}) and map them to their
corresponding ecliptic coordinates $(D, \beta,\lambda)$.

The physical properties of the WD-WD population (${\cal M}_c \equiv
{(m_1 m_2)}^{3/ 5}/{(m_1 + m_2)}^{1/5}$, with $m_1$, $m_2$ being the
masses of the two stars, and $\omega_s = 2 \pi f_s = 4 \pi/{\rm orbital
  period}$) are taken from the binary population synthesis simulation
discussed in \cite{NYP04}. For details on this simulation we refer the
reader to \cite{NYP04}, and for earlier work to
\cite{HBW90,EIS87,HB90,HB00,BH97,NYP01}. The basic ingredient for
these simulations is an approximate binary evolution code. A
representation of the complete Galactic population of binaries is
produced by evolving a large (typically $10^6$) number of binaries
from their formation to the current time, where the distributions of
the masses and separations of the initial binaries are estimated from
the observed properties of local binaries.  This initial-to-final
parameter mapping is then convolved with an estimate of the binary
formation rate in the history of the Galaxy to obtain the total
Galactic population of binaries at the present time.  From these the
binaries of interest can then be selected.  In principle this
technique is very powerful, although the results can be limited by the
limited knowledge we have on many aspects of binary evolution. For
WD-WD binaries, the situation is better than for many other
populations, since the observed population of WD-WD binaries allows us
to gauge the models (e.g. \cite{NYP01}).

We also include the population of semi-detached WD-WD binaries
(usually referred to as AM CVn systems) that are discussed in detail
in \cite{NYP04}.  In these binaries one white dwarf transfers its
outer layers onto a companion white dwarf. Due to the redistribution
of mass in the system, the orbital period of these binaries increases
in time, even though the angular momentum of the binary orbit still
decreases due to gravitational wave losses. The formation of these
systems is very uncertain, mainly due to questions concerning the
stability of the mass transfer (e.g. \cite{MNS04})

From the models of the Galactic population of the detached WD-WD
binaries and AM CVn systems two dimensional histograms were created,
giving the expected number of both WD-WD binaries and AM CVn systems
currently present in the Galaxy as function of the $log$ of the GW
radiation frequency, $f_s (= \omega_s/2 \pi)$ and chirp mass,
$\mathcal{M}_c$.  In the case of the detached WD-WD binaries, the
($\log f_s, \mathcal{M}_c)$) space was defined over the set
$\mathcal{M}_c \in (0, 1.5]$, $\log f_s \in [-6, -1]$, and contained
$30 \ \times \ 50$ grid points, while in the case of the AM CVn
systems the region is intrinsically smaller, $\mathcal{M}_c \in (0,
1.2]$, $\log f_s \in [-4, -1.5]$, containing only $24 \ \times \ 25$
grid points.

\begin{figure}
\includegraphics[width=10cm]{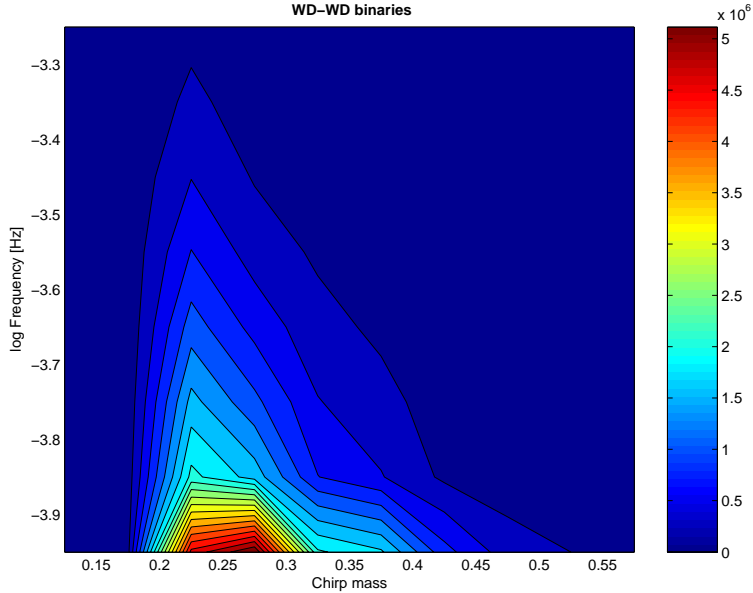}
\caption{The distribution of detached white-dwarf --- white-dwarf
  binaries in our galaxy as a function of the gravitational wave
  frequency, $f_s$, and chirp mass, ${\mathcal M}_c$.}
\label{fig:wdwd_nel1}
\end{figure}

\begin{figure}
\includegraphics[width=10cm]{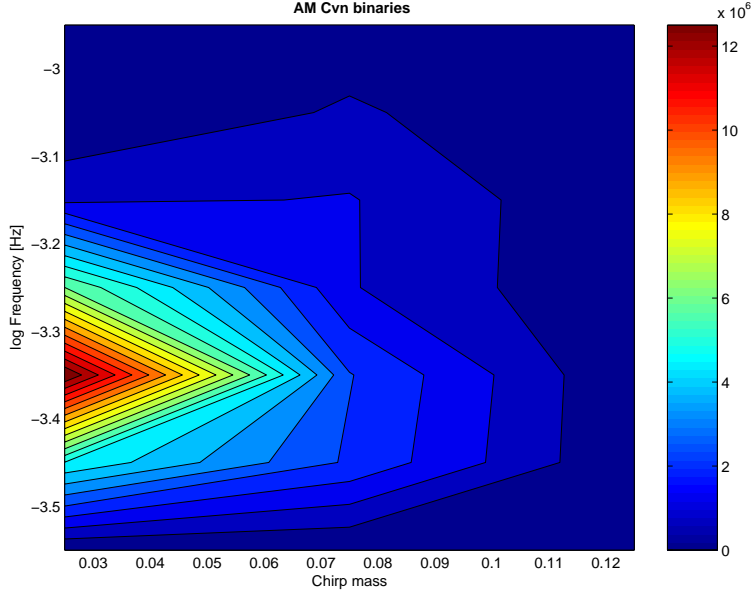}
\caption{The distribution of AM CVn binary systems in our galaxy as a
    function of the gravitational wave frequency, $f_s$, and chirp
    mass, ${\mathcal M}_c$.}
\label{fig:amcvn_nel1}
\end{figure}

Figure (\ref{fig:wdwd_nel1}) shows the distribution of the number of
detached WD-WD binaries as a function of the chirp mass and signal
frequency in the form of a contour plot. This distribution reaches its
maximum within the LISA frequency band when the chirp mass is equal to
$\simeq 0.25 \ {\cal M}_\odot$, and it monotonically decreases as a
function of the signal frequency.  The distribution of the number of
AM CVn systems has instead a rather different shape, as shown by the
contour plot given in Figure (\ref{fig:amcvn_nel1}). The region of the
($\mathcal{M}_c, \log f_s$) space over which the distribution is
non-zero is equal to ($0, 0.07$) $\times$ ($-3.4, -2.2$), and it
reaches its maximum at the point ($0.03, -3.35$).


\section{Simulation of the background signal in the LISA data}
\label{Simu}

In order to simulate the LISA $X$ response to the population of WD-WD
binaries derived in Section \ref{WD_POP} one needs to coherently add
the LISA response to each individual signal. Although this could
naturally be done in the time domain, the actual CPU time required to
successfully perform such a simulation would be unacceptably long. The
generation in the time domain of one year of $X^{\rm GW} (t)$ response
to a single signal sampled at a rate of $16$ seconds would require
about $1$ second with an optimized C++ code running on a Pentium IV
$3.2$ GHz processor.  Since the number of signals from the background
is of the order $10^{8}$, it is clear that a different algorithm is
needed for simulating the background in the LISA data within a
reasonable amount of time. We were able to derive and implement
numerically an analytic formula of the Fourier transform of each
binary signal, which has allowed us to reduce the computational time
by almost a factor $100$.  Furthermore, we have run our code on the
Jet Propulsion Laboratory (JPL) supercomputer system, which includes
$64$ Intel Itanium2 processors each with a clock speed of 900 MHz.

\subsection{The Fourier transform of a binary signal}
\label{X_FOURIER}

The expression of the Fourier transform of the TDI response
$X^\mathrm{GW}$ to a single binary signal (Eqs. (\ref{X}, \ref{A},
\ref{phi}), cannot be written (to our knowledge) in closed analytic
form.  However, by using the LWL expansion of the $X^{\rm
  GW}$-response, it is possible to obtain a closed-form expression of
its Fourier transform.  Since the WD-WD binary background has a
natural frequency cut-off that is between $1$ and $2$ millihertz, the
LWL expansion of the $X^{\rm GW}$ response (Eq \ref{Xn}), truncated at
$n=2$, can be used for accurately representing the gravitational wave
response of each binary signal, as discussed in Section \ref{LISA_RES}.

In order to derive the Fourier transform of $X^{\rm GW}_{(2)} (t)$, we
use the following expansion of the function $e^{- i \phi (t)}$ in terms of
the Bessel functions of the first kind, $J_q$, \cite{AS72}
\begin{equation}
e^{-i \ \phi (t)} =  \sum_{q=-\infty}^{\infty} J_q (\omega_s R \cos\beta) \ e^{- i[\omega_s t \ + \ q \
  (\Omega t \ + \ \eta_0 \ - \ \lambda \ + \  \frac{\pi}{2})]} \ .
\label{Bessel}
\end{equation}
Since the Bessel functions $|J_q(\omega_s R \cos\beta)|$ are much
smaller than unity when $|q| >> |\omega_s R \cos\beta|$, the expansion
given by equation (\ref{Bessel}) can be truncated at a finite index
$Q$, providing an accurate numerical estimation of the function $e^{-i
  \ \phi (t)}$.  These considerations allow us to write the following
expression of the Fourier transform of $X^{\rm GW}_{(n)} (t)$
\begin{eqnarray} 
\widetilde{X}^{\rm GW}_{(n)} (\omega) & = & \mathcal{F} 
\left( \Re \sum_{k=0}^n \sum_{q= - Q}^{Q}  A^{(k)} (t) \ x^{k+2} \ J_q(\omega_s R \cos\beta) \ e^{- i[\omega_s t \ + \ q \
  (\Omega t \ + \ \eta_0 \ - \ \lambda \ + \  \frac{\pi}{2})]} \right) 
\nonumber
\\ 
& = &  \pi \sum_{q=- Q}^{Q}  J_q(\omega_s R \cos\beta) \sum_{k=0}^n x^{k+2} \left\{
\mathcal{F}\left(\Re [A^{(k)} (t)] \right) * \left[\delta(\omega +
\omega_s + q \Omega)  e^{i q (-\eta_0+\lambda-\pi/2)} \right. \right.
\nonumber
\\
&& + \left. \left. \delta(-\omega+\omega_s + q \Omega) e^{i q(
      \eta_0- \lambda+\pi/2)} \right] +
i \mathcal{F} \left(\Im [A^{(k)} (t)] \right) * \left[\delta(\omega +
\omega_s + q \Omega) e^{i q (-\eta_0+\lambda-\pi/2)} \right. \right.
\nonumber
\\
&& - \left. \left. \delta(-\omega+\omega_s + q \Omega) e^{i q( \eta_0-
    \lambda+\pi/2)} \right] \right\} \ ,
\label{XF}
\end{eqnarray}
where $\mathcal{F}$ is the Fourier transform operator, the symbol $*$
between two expressions means their convolution, $\omega$ is the
Fourier angular frequency, and $A^{(k)} (t)$ are defined in Eq. (\ref{Xn})
and given in Eq. (\ref{An}) with $k= 0, 1, 2$.

As an example application of this general formula for the Fourier
transform of the $X^{\rm GW}_{(n)} (t)$ response, let us apply it to
the lowest order LWL expansion ($n=0$)
\begin{eqnarray} 
\widetilde{X}^{\rm GW}_{(0)} (\omega) & = & 4 \pi \sum_{q = - Q}^{Q}
J_q(\omega_s R \cos\beta) x^{2} \left\{ 
\left[a_1 (\tilde{u}_2(\omega)-\tilde{u}_3(\omega)) + a_2
  (\tilde{v}_2(\omega)-\tilde{v}_3(\omega)) \right] \right.
\nonumber
\\
&& \left. * \left[\delta(\omega + \omega_s + q \Omega) 
e^{i q (-\eta_0+\lambda-\pi/2)} + \delta(-\omega+\omega_s + q \Omega) e^{i q( \eta_0-
      \lambda+\pi/2)} \right] \right.
\nonumber
\\
&& \left. + i \left[a_3 (\tilde{u}_2(\omega)-\tilde{u}_3(\omega)) + a_4
  (\tilde{v}_2(\omega)-\tilde{v}_3(\omega))\right] \right.
\nonumber
\\
&& \left. * \left[\delta(\omega + \omega_s + q \Omega) e^{i q
    (-\eta_0+\lambda-\pi/2)} - \delta(-\omega+\omega_s + q \Omega) e^{i q( \eta_0-
  \lambda+\pi/2)} \right] \right\} \ .
\label{XF0}
\end{eqnarray}
Since the Fourier transforms of $u$ and $v$ are both linear
combinations of nine Dirac delta functions centered on the frequencies
$\pm l \ \Omega \ , \ l=0, 1, 2, 3, 4$ (see equations (27--30) in
reference \cite{KTV04} for the expressions of $u$ and $v$), it follows
that $\tilde{X}^{\rm GW}_{(0)} (\omega)$ is also a linear combination
of Dirac delta functions. In particular, in the limit of negligible
Doppler modulation, the resulting expression (\ref{XF0}) reduces as
expected to that of a purely amplitude modulated sinusoidal signal
with central frequency equal to $\omega_s$ and upper and lower
band-limits given by $\omega_s + 4 \Omega$ and $\omega_s - 4 \Omega$
respectively \cite{GP97}.

The actual expression of the Fourier transform we implemented in our
simulation of the WD-WD background used Eq. (\ref{XF}) with $n=2$, and
maximum value of the index of the Bessel expansion, $Q$, equal to $
|\omega_s R \cos \beta| + 20$ in order to make negligible the error
associated with the truncation of the expansion itself.

One extra mathematical detail that we need to include is that the
Fourier transform of $X^{\rm GW}_{(n)} (t)$ is performed over a finite
integration time, $T$, while the expression in Eq. (\ref{XF})
corresponds to an infinite-time Fourier transform. In order to account
for this discrepancy we convolved the analytic Fourier transform of
the signal given in equation (\ref{XF}) with the Fourier transform of
a window function with an integration time $T$.  To avoid leakage
introduced by using a simple rectangular window, we have used instead
the Nuttall's modified Blackman-Harris window \cite{NBH81}. Although
this window is characterized by having the main lobe of its Fourier
transform slightly wider than that of the rectangular window, the
maximum of its side-lobes are about four orders of magnitude lower
than those of the rectangular window, reducing leakage significantly.
The expression of its Fourier transform, ${\widetilde {W}} (\omega)$,
is equal to
\begin{eqnarray}
{\widetilde {W}} (\omega) & = & n_0
[Sinc(\omega T) + i Cosc(\omega T)]
\label{WF}
\\
& - & n_1 \ 
[Sinc(\omega T + 2 \pi) + Sinc(\omega T - 2 \pi) 
+ i \ (Cosc(\omega T + 2 \pi) + Cosc(\omega T - 2 \pi))] 
\nonumber
\\
& - & n_2 \ [Sinc(\omega T + 4 \pi) + Sinc(\omega T - 4
\pi) 
+ i \ (Cosc(\omega T + 4 \pi) + Cosc(\omega T - 4 \pi))]
\nonumber
\\
& - & n_3 \ [Sinc(\omega T + 6 \pi) + Sinc(\omega T - 6 \pi)
+ i \ (Cosc(\omega T + 6 \pi) + Cosc(\omega T - 6 \pi))] \ ,
\nonumber
\end{eqnarray}
where the functions $Sinc (.)$ and $Cosc (.)$ are defined as follows
\begin{equation}
Sinc(.) \equiv \frac{\sin(.)}{.} \ \ , \ \ Cosc(.) \equiv
\frac{\cos(.) - 1}{.} \ ,
\end{equation}
and  the coefficients $n_r \ , \ r=0, 1, 2, 3$
have the following numerical values
\begin{equation}
n_0 = 0.3635819 \ \ , \ \ 
n_1 = 0.24458875 \ \ , \ \ 
n_2 = 0.06829975\ \ , \ \ 
n_3 = 0.00532055\ .
\end{equation}

\subsection{Generation of the signal parameters}

We used the distributions given in Section \ref{WD_POP} to randomly
generate the parameters $\phi_o$, $\iota$, $\psi$, $D$, $\beta,
\lambda$, while the values of the chirp mass, ${\cal M}_c$, and the
logarithm of the frequency of the signal, $\log(f_s)$, were obtained
by further processing the numeric distribution function (given in
Section \ref{WD_POP}) of the number of sources. To derive the
distribution function for the variables (${\cal M}_c, \log (f_s)$)
within each grid-rectangle of our numerical distribution we proceeded
in the following way \cite{ValPriv04}.  Let us consider the number of
sources $N (x_1, x_2)$ as a function of two coordinates ($x_1, x_2$)
of a point in the $({\cal M}_c, \log (f_s))$ plane within a specified
grid-rectangle of the numerical distribution.  This function can be
approximated there by the following quadratic polynomial
\begin{equation}
N (x_1,x_2)= n_{00} (1-x_1) (1-x_2) + n_{11} x_1 x_2 + 
n_{01} (1-x_1) x_2 + n_{10} x_1 (1-x_2) \ ,
\end{equation}
where $n_{00}$, $n_{11}$, $n_{01}, n_{10}$ are equal to the number of
signals at the ``corners'' of the considered grid-rectangle and are
obtained by interpolation; ($x_1, x_2$) are therefore two real numbers
defined in the range $[0, 1]$.

If we integrate out the $x_2$-dependence in $N(x_1,x_2)$ we obtain
\begin{equation}
N_{x_2} (x_1) = \int_0^1 N (x_1,x_2) dx_2 = 
\frac{(-n_{00}+n_{11}-n_{01}+n_{10}) x_1 + n_{00} + n_{01} }{2} \ ,
\end{equation}
which defines the total number of sources within that grid-rectangle
having chirp mass equal to $x_1$. In order to derive the probability
distribution function of $x_1$ within that grid-rectangle we can
define the following mapping between a uniformly distributed random
variable, say $z_1$, and the random variable $x_1$ 
\begin{equation}
z_1 =  \frac{\int_0^{x_1} N_{x_2}(x_1') dx_1'}{\int_0^1 N_{x_2}(x_1') dx_1'} = 
\frac{(-n_{00}+n_{11}-n_{01}+n_{10}) x_1^2 + 2 (n_{00} + n_{01}) x_1 }
{n_{00}+n_{11}+n_{01}+n_{10}} \ .
\end{equation}
By solving the above non-linear equation for
every uniformly sampled $z_1$ we obtain
\begin{equation}
x_1 = \frac{n_{00}+n_{01} - \sqrt{n_{00}^2+2 n_{00} n_{01}+n_{01}^2 + 
(- n_{00}^2 - 2 n_{00} n_{01} + n_{11}^2 + 2 n_{11} n_{10}-n_{01}^2+
n_{10}^2) z_1 }}{n_{00}-n_{11}+n_{01}-n_{10}} \ ,
\label{x1}
\end{equation}
where the branch ``-'' has been chosen such that $x_1$ remains in the range
$[0, 1]$.  If $n_{00}-n_{11}+n_{01}-n_{10} = 0$ then equation
 (\ref{x1}) is no longer valid, and we have instead $x_1 = z_1$. 

A similar procedure can be implemented for calculating $x_2$. By
integrating $N (x_1, x_2')$ with respect to $x_2'$ over the
range ($0, x_2$), we can establish the following relationship between
another uniformly distributed random variable, say $z_2$, and $x_2$
\begin{equation}
z_2 = \frac{\int_0^{x_2} N(x_1,x_2') dx_2'}{\int_0^{1} N(x_1,x_2') dx_2'} = 
\frac{[(n_{00}+n_{11}-n_{01}-n_{10}) x_1-n_{00}+n_{01}] x_2^2+
2 [(-n_{00}+n_{10}) x_1+n_{00}] x_2}{(-n_{00}+ n_{11}- n_{01} + n_{10})
  x_1+n_{00}+n_{01}} \ .
\end{equation}
After some simple algebra we can finally solve for $x_2$ in terms of
$x_1$ (itself a function of the uniformly distributed random variable
$z_1$) and $z_2$
\begin{equation}
x_2 = \frac{n_{00}(x_1 -1) -n_{10} x_1 +
\sqrt{F(x_1) z_2 
+(n_{00}^2-2 n_{00} n_{10}+n_{10}^2) x_1^2+ 2 (-n_{00}^2+n_{00}
n_{10}) x_1+n_{00}^2}}
{(n_{11}-n_{01}-n_{10}+n_{00}) x_1 - n_{00}+n_{01}} \ ,
\end{equation}
where now we have chosen the ``+'' branch so $x_2 \in [0, 1]$ range,
and the function $F(x_1)$ is equal to
\begin{equation}
F(x_1) \equiv (n_{11}^2 + n_{01}^2 - n_{00}^2 - 2 n_{01} n_{11} + 2 n_{00}
  n_{10} - n_{10}^2) x_1^2 +2 (-n_{00} n_{10} +
  n_{00}^2-n_{01}^2+n_{01} n_{11}) x_1-n_{00}^2+n_{01}^2 \ .
\end{equation}
Note that the equation for $x_2$ above is no longer valid when
$(n_{11}-n_{01}-n_{10}+n_{00}) x_1 - n_{00}+n_{01} = 0$, in which case
$x_2 = z_2$.  Once $x_1$ and $x_2$ are calculated, they can be
converted into the physical parameters ${\cal M}_c$ and $\omega_s$
according to the following relationships
\begin{eqnarray}
{\cal M}_c & = & {{\cal M}_c}_{00} +  x_1 \ \Delta_{{\cal M}_c} \\
\omega_s & = & 2 \pi 10^{\log ({f_s}_{00}) + x_2 \ \Delta_{\log (f_s)}}
\end{eqnarray}
where (${{\cal M}_c}_{00}, \log ({f_s}_{00})$) are the coordinates of
the ``lower-left-hand-corner'' of the considered grid-rectangle, and
$\Delta_{{\cal M}_c}$, $\Delta_{\log ({f_s})}$ are the lengths of the
sides of the grid-rectangle.

\subsection{Results of the Numerical Simulation}
\label{NumSim}

\begin{figure}
\includegraphics[width=6in]{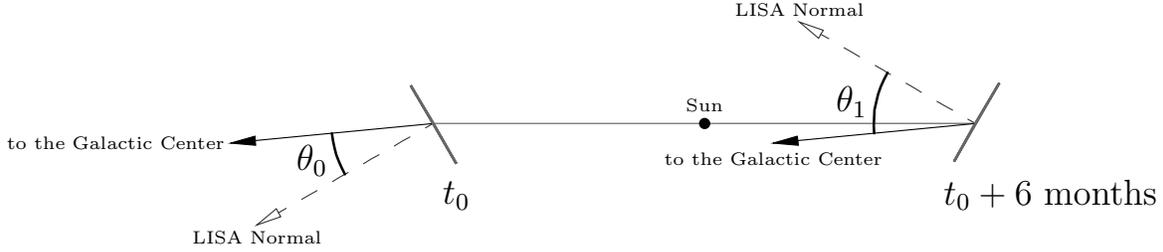}
\caption{Two snapshots of LISA along its trajectory as recorded six
  months apart by an observer in the ecliptic plane. They correspond
  to the times when the LISA response to the background achieves a local
  maximum.  The magnitudes of these maxima are not equal due to the
  relative disposition of the ecliptic plane with respect to the
  galactic plane. At time $t_0$ the angle $\theta_0$ between the
  normal to the plane of LISA and a vector pointing to the galactic
  center is equal to $24.47^\circ$.  Six months later, when the LISA
  response is also at a maximum, the angle $\theta_1$ is equal to
  $35.53^\circ$ which results in a smaller maximum.}
\label{Galaxy_Ecliptic}
\end{figure}

\begin{figure}
\includegraphics[width=6in]{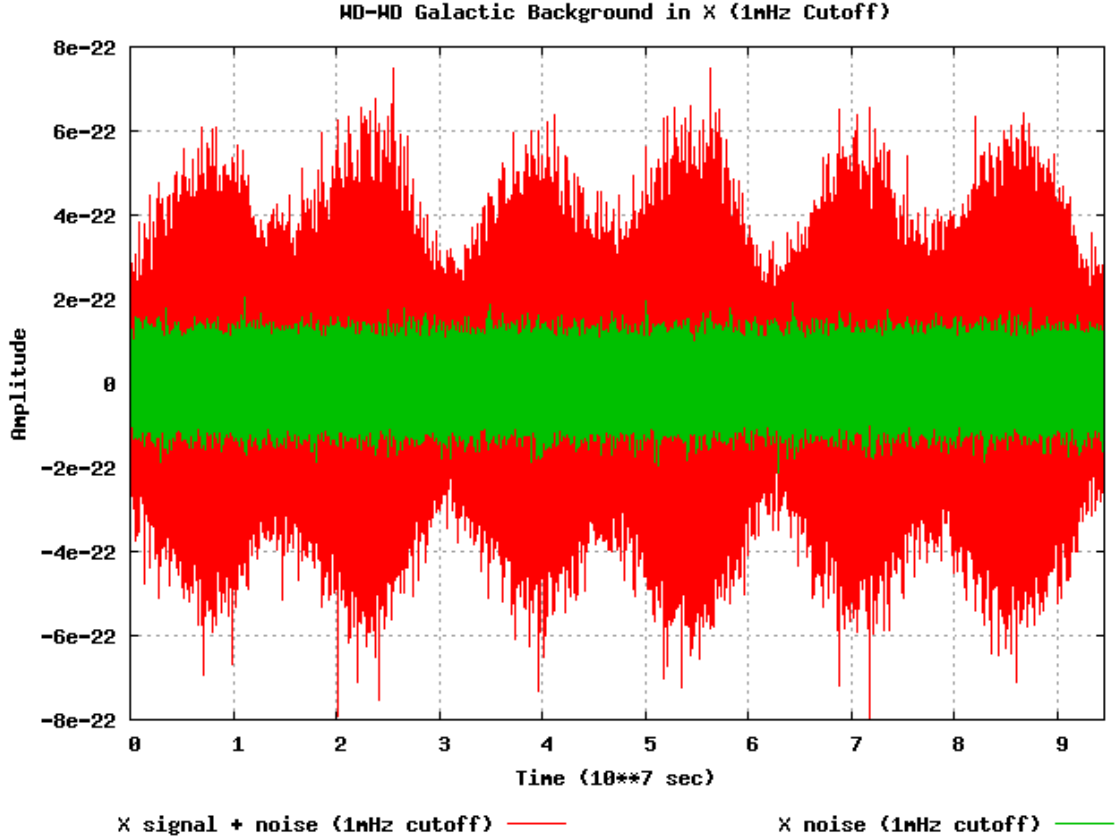}
\caption{Three years of simulated TDI $X$ response to
  the WD-WD Galactic background signal. The time series of the LISA
  instrumental noise is displayed for comparison.}
\label{time_results}
\end{figure}

The expression for the finite-time Fourier transform of each WD-WD
signal in the $\widetilde{X}^{\rm GW}$ response given in Section
\ref{X_FOURIER} allows us to coherently add in the Fourier domain all
the signals radiated by the WD-WD galactic binary population described
in Section \ref{WD_POP}.  After inverse Fourier transforming the
synthesized response and removing the window from it, we finally
obtain the time-domain representation of the background as it will be
seen in the LISA TDI combination $X$.  This is shown in Figure
(\ref{time_results}), where we plot three years worth of simulated
$X^{\rm GW} (t)$, and include the LISA noise \cite{PPA98}. The
one-year periodicity induced by the motion of LISA around the Sun is
clearly noticeable.  One other interesting feature shown by Figure
(\ref{time_results}) is that the amplitude response reaches absolute
minima when the Sun-LISA direction is roughly oriented towards the
Autumn equinox, while the absolute maxima take place when the Sun-LISA
direction is oriented roughly towards the Galactic center
\cite{Seto04}. This fact can easily be understood by looking at Figure
(\ref{Galaxy_Ecliptic}). Since the ecliptic plane is not parallel to
the galactic plane, and because our own solar system is about $8.5 \ 
{\rm kpc}$ away from the galactic center (where most of the of WD-WD
binaries are concentrated ) it follows that the LISA $X^{\rm GW}$
response does not have a six-months periodicity.

Note also that, for a time period of about $2$ months, the absolute
minima reached by the amplitude of the LISA response to the WD-WD
background is only a factor less than $2$ larger than the level of the
instrumental noise.  This implies that during these observation times
LISA should be able to search for other sources of gravitational
radiation that are not located in the galactic plane. This might turn
out to be the easiest way to mitigate the detrimental effects of the
WD-WD background when searching for other sources of gravitational
radiation.  We will quantitatively analyze in a follow up work how to
take advantage of this observation in order to optimally search, during
these time periods, for sources that are off the galactic plane.

\begin{figure}
\includegraphics[width=6in]{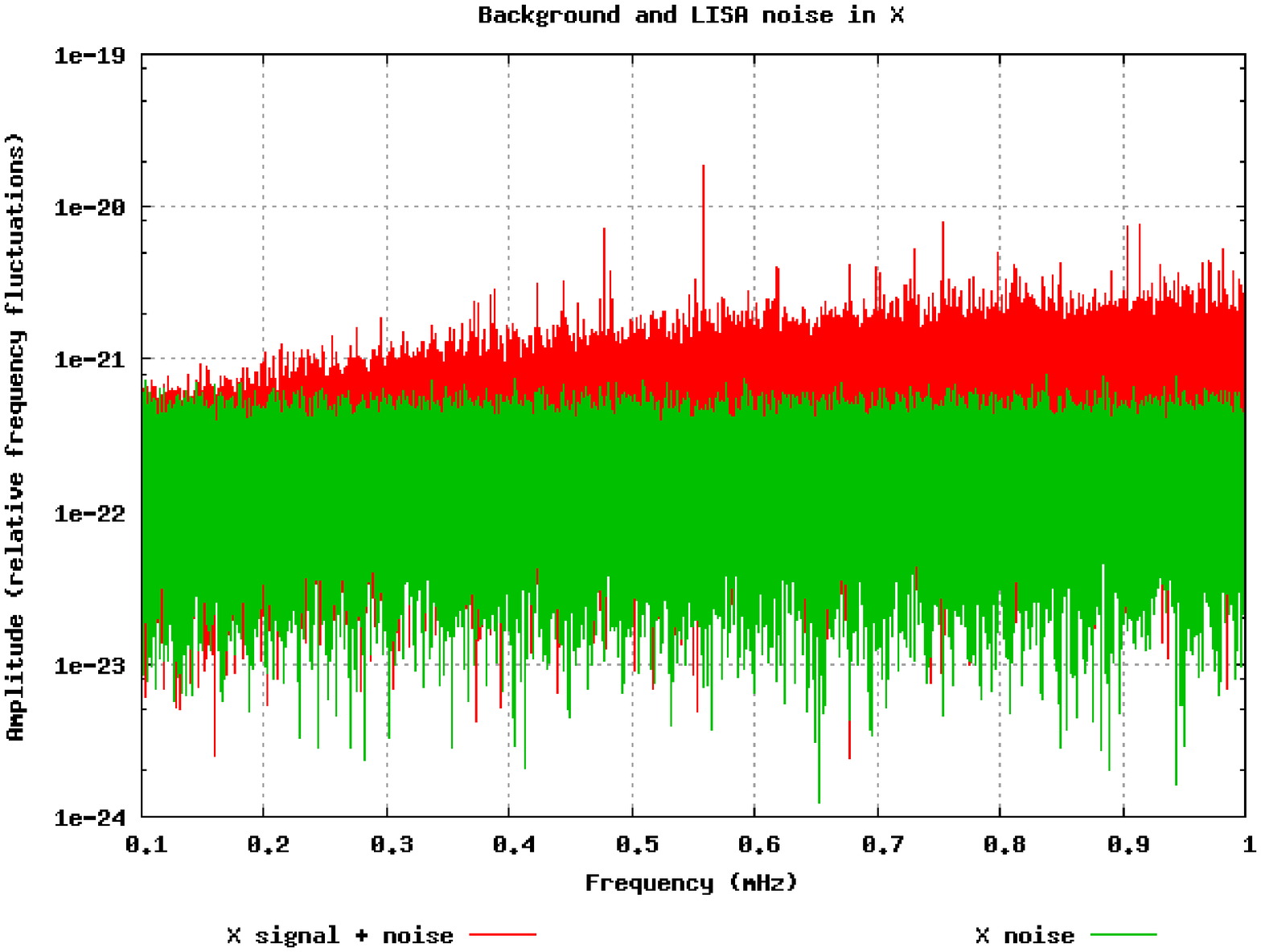}
\caption{The amplitude of the Fourier transform of the WD-WD Galactic
  background gravitational wave signal and of the LISA instrumental
  noise entering into the TDI combination $X$.}
\label{Spectrum}
\end{figure}

In Figure (\ref{Spectrum}) we plot, as functions of the Fourier
frequency, $f$, the windowed Fourier powers of both the signal and the
noise entering into the TDI $X$ combination.  Note that in the region
of the LISA band below $0.2$ millihertz the power of the WD-WD
background is smaller than that of the instrumental noise.

\section{Cyclostationary processes}
\label{sec:cyclo}

The results of our simulation (Figure (\ref{time_results})) indicate
that the LISA $X^{\rm GW}$ response to the background can be regarded,
in a statistical sense, as a periodic function of time. This is
consequence of the deterministic (and periodic) motion of the LISA
array around the Sun. Since its autocorrelation function will also be
a periodic function of period one year, it follows that any LISA
response to the WD-WD background should no longer be treated as a
stationary random process but rather as a periodically correlated
random process. These kind of processes have been studied for many
years, and are usually referred to as cyclostationary random processes
(see \cite{H89} for a comprehensive overview of the subject and for more
references). In what follows we will briefly summarize the properties
of cyclostationary processes that are relevant to our problem.

A continuous stochastic process ${\cal X} (t)$ having finite second order
moments is said to be {\em cyclostationary\/} with period $T$ if the
following expectation values
\begin{eqnarray}
E[{\cal X} (t)] &=& m(t) = m(t + T) , \\
E[{\cal X} (t') {\cal X} (t)] &=& C(t',t) = C(t' + T,t + T)
\end{eqnarray}
are periodic functions of period $T$, for every $(t',t) \in {\bf R}
\times {\bf R}$. For simplicity from now on we will assume $m(t) = 0$.

If ${\cal X} (t)$ is cyclostationary, then the function $B(t,\tau) \equiv C(t
+ \tau,t)$ for a given $\tau \in {\bf R}$ is periodic with period $T$,
and it can be represented by the following Fourier series
\begin{equation}
B(t,\tau) = \sum_{r = -\infty}^{\infty}B_r(\tau) e^{i
2\pi\frac{r t}{T}} \ ,
\end{equation}
where the functions $B_r(\tau)$ are given by
\begin{equation}
B_r(\tau) = \frac{1}{T}\int^T_0B(t,\tau) e^{- i 2\pi r\frac{
    t}{T}} \ dt \ .
\label{eq:FB}
\end{equation}
The Fourier transforms $g_r(f)$ of $B_r(\tau)$ are the so called
``cyclic spectra'' of the cyclostationary process ${\cal X} (t)$  \cite{H89}
\begin{equation}
g_r(f) = \int_{-\infty}^{\infty}B_r(\tau)e^{-i 2\pi f \tau } \ d
\tau \ .
\end{equation}
If a cyclostationary process is real, the following relationships
between the cyclic spectra hold
\begin{eqnarray}
B_{-r}(\tau) & = & B^*_r(\tau) \ ,
\label{eq:ksym1}
\\
g_{-r}(-f) & = &  g^*_r(f) \ ,
\label{eq:ksym2}
\end{eqnarray}
where the symbol $^*$ means complex conjugation.  This implies that,
for a real cyclostationary process, the cyclic spectra with $r \geq 0$
contain all the information needed to characterize the process itself.

The function $\sigma^2(\tau) \equiv B(0,\tau)$ is the variance of the
cyclostationary process ${\cal X} (t)$, and it can be written as a Fourier
decomposition as a consequence of Eq. (\ref{eq:FB})
\begin{equation}
\sigma^2(\tau) = \sum_{r=-\infty}^{\infty}H_r e^{ i
2\pi\frac{r \tau}{T}},
\end{equation}
where $H_r \equiv B_r(0)$ are harmonics of the variance $\sigma^2$.
From Eq. (\ref{eq:ksym1}) it follows that $H_{-r} =
{H}^*_r$.

For a discrete, finite, real time series ${\cal X}_t$, $t =
1,\ldots, {\cal N}$ we can estimate the cyclic spectra by generalizing
standard methods of spectrum estimation used with stationary
processes. Assuming again the mean value of the time series ${\cal
  X}_t$ to be zero, the cyclic autocorrelation sequences are defined
as
\begin{equation}
s_l^r = \frac{1}{\cal N}\sum_{t=1}^{{\cal N}-|l|}{\cal X}_t {\cal X}_{t+|l|}
e^{-\frac{i 2\pi r (t-1)}{T}} \ . 
\label{eq:cycorr}
\end{equation}
It has been shown \cite{H89} that the cyclic autocorrelations are
asymptotically (i.e.\ for $N \rightarrow \infty$) unbiased estimators
of the functions $B_r(\tau)$. The Fourier transforms of the cyclic
autocorrelation sequences $s_l^r$ are estimators of the cyclic spectra
$g_r(f)$. These estimators are asymptotically unbiased, and are called
``inconsistent estimators'' of the cyclic spectra, i.e. their
variances do not tend to zero asymptotically.  In the case of Gaussian
processes \cite{H89} consistent estimators can be obtained by first
applying a lag window to the cyclic autocorrelation and then perform
a Fourier transform.  This procedure represents a generalization of
the well-known technique for estimating the spectra of stationary
random processes \cite{PW93}.

An alternative procedure for identifying consistent estimators of the
cyclic spectra is to first take the Fourier transform,
$\tilde{{\cal X}}(f)$, of the time series ${\cal X} (t)$
\begin{equation}
\tilde{{\cal X}}(f) = \sum_{t = 1}^{\cal N}  {\cal X}_t e^{-i 2\pi f (t - 1)}
\end{equation}
and then estimate the cyclic periodograms $g_r(f)$
\begin{equation}
g_r(f) = \frac{\tilde{{\cal X}}(f)\tilde{{\cal X}}^*(f - \frac{2\pi
    r}{T})}{\cal N} \ .
\end{equation}
By finally smoothing the cyclic periodograms, consistent estimators of
the spectra $g_r(f)$ are then obtained.  The estimators of the
harmonics $H_r$ of the variance $\sigma^2$ of a cyclostationary
random process can be obtained by first forming a sample variance of
the time series ${\cal X}_t$. The sample variance is obtained by
dividing the time series ${\cal X}_t$ into contiguous segments of
length $\tau_0$ such that $\tau_0$ is much smaller than the period $T$ of
the cyclostationary process, and by calculating the variance
$\sigma^2_I$ over each segment.  Estimators of the harmonics are
obtained either by Fourier analyzing the series $\sigma^2_I$ or by
making a least square fit to $\sigma^2_I$ with the appropriate number
of harmonics. Note that the definitions of (i) zero order ($r = 0$)
cyclic autocorrelation, (ii) periodogram, and (iii) zero order
harmonic of the variance, coincide with those usually adopted for
stationary random processes.  Thus, even though a cyclostationary time
series is not stationary, the ordinary spectral analysis can be used
for obtaining the zero order spectra.  Note, however, that
cyclostationary random processes provide more spectral information
about the time series they are associated with due to the existence of
cyclic spectra with $r > 0$.

As an important and practical application, let us consider a time
series $y_t$ consisting of the sum of a stationary random process,
$n_t$, and a cyclostationary one ${\cal X}_t$ (i.e. $y_t = n_t + {\cal
  X}_t$).  Let the variance of the stationary time series $n_t$ be
$\nu^2$ and its spectral density be $\mathcal{E}(f)$.  It is
easy to see that the resulting process is also cyclostationary. If the
two processes are uncorrelated, then the zero order harmonic
$\Sigma^2_0$ of the variance of the combined processes is equal to
\begin{equation}
\Sigma^2_0 = \nu^2 + \sigma^2_0 \ ,
\end{equation}
and the zero order spectrum, $G_0(f)$, of $y_t$ is
\begin{equation}
G_0(f) =  \mathcal{E}(f)  +  g_0(f) \ .
\end{equation}
The harmonics of the variance as well as the cyclic spectra of $y_t$
with $r > 0$ coincide instead with those of ${\cal X}_t$.  In other words,
the harmonics of the variance and the cyclic spectra of the process
$y_t$ with $r > 0$ contain information only about the cyclostationary
process ${\cal X}_t$, and are not ``contaminated'' by the stationary process
$n_t$. 

\section{Analytic study of the background signal}
\label{sec:cycloa}

In the case of the ensemble of $N$ WD-WD binaries, the total signal
$s(t)$ is given by the following sum
\begin{equation}
s(t) = \sum^N_{i=1} X^{\rm GW} (t; {\bf \Lambda}_i) \ , 
\end{equation}
where ${\bf \Lambda}$ represents the set
($\phi_o,\iota,\psi,D,\beta,\lambda,{\cal M}_c, \omega_s$) of $8$
parameters characterizing a GW signal. Since $N$ is large, we can
expect the parameters of the signals to be randomly distributed and
regard the signal $s(t)$ itself as a random process. Its mean, $m(t)$,
and its autocorrelation function, $C(t',t)$ can then be calculated by
assuming the probability distribution of the vector $\bf \Lambda$,
$P({\bf \Lambda})$, to be the product of the five probability
distributions, $P_1 (\phi_o)$, $ P_2 (\iota)$, $ P_3 (\psi)$, $ P_4
(D, \beta, \lambda)$, and $ P_5({\cal M}_c, \omega_s)$ (as we did in
our numerical simulation of the WD-WD background in Section
\ref{WD_POP}).  By assuming the angles $\phi_o$ and $\psi$ to be
uniformly distributed in the interval $[0, 2\pi)$, and $\cos\iota$ to
be uniformly distributed in the interval $[-1,1]$, we can then perform
the integrals over the angles $\phi_o$, $\psi$, and $\iota$
analytically and obtain the following expressions
\begin{eqnarray}
m(t) & = & N \int_V  X^\mathrm{GW}(t) P({\bf \Lambda}) d {\bf \Lambda}
\nonumber
\\
& = & \frac{N}{8 \pi^2}\int_0^{2\pi}d\phi_o \int_0^{2\pi}d\psi 
\int_{-1}^1d\cos\iota
\int_{V_4}\int_{V_5} X^\mathrm{GW}(t) P_4 P_5 dV_4 dV_5 = 0 \ , 
\end{eqnarray}
\begin{eqnarray}
C(t',t) & = & N \int_V  X^\mathrm{GW}(t') X^\mathrm{GW}(t) 
 P({\bf \Lambda}) d {\bf \Lambda}
\nonumber
\\
& = & \frac{N}{16 \pi^2}\int_0^{2\pi}d\phi_o 
\int_0^{2\pi}d\psi \int_{-1}^1d\cos\iota
\int_{V_4}\int_{V_5} \Re [A(x, t') A^* (x, t) 
e^{i[\phi(t) - \phi(t')]}] P_4 P_5 dV_4
dV_5 \ .
\end{eqnarray}
Note that the mean value $m (t)$ is equal to $0$ as a consequence of
averaging the antenna response over the polarization angle $\psi$.

In order to gain an analytic insight about the statistical properties
of the autocorrelation function $C(t', t)$, in what follows we will
adopt the zero-order long-wavelength approximation of the LISA
response $X^{\rm GW} (t)$ obtained by fixing $n=0$ in (\ref{Xn}) and
using the expression for the complex amplitude $A^{(0)}$ given in
equation (\ref{An}).  After some long but straightforward algebra, the
autocorrelation function, $C(t', t)$, can be written in the following
form
\begin{eqnarray}
C(t',t) & = & \frac{16}{5} N \int_{V_4}\int_{V_5} x^4  h_o^2  
[u_2 (t') u_2 (t) + v_2 (t') v_2 (t) + u_3 (t') u_3 (t) + v_3 (t') v_3
(t) 
\nonumber
\\
& - &
u_2 (t') u_3 (t) - v_2 (t') v_3 (t) -
u_3 (t') u_2 (t) - v_3 (t') v_2 (t)]
\cos[\phi(t') - \phi(t)] P_4 P_5 \ dV_4 \ dV_5 \ ,
\end{eqnarray}
where $x = \omega_s L$, and $h_o = \frac{4 {{\cal M}_c}^{5/3}}{D}
\left[\frac{\omega_s}{2}\right]^{2/3}$ (units in which the
gravitational constant, $G$, and the speed of light, $c$, are equal to
$1$).  For frequencies less than $1$ mHz the Doppler modulation in the
phase $\phi(t)$ can be neglected making $\phi(t) \simeq \omega_s \,
t$. If we now introduce a new time variable $\tau = t' - t$ and define
$B(t,\tau) \equiv C(t + \tau,t)$, we have
\begin{equation}
B(t,\tau) = 
\int_{0}^{\infty}
{\cal P}(\omega_s) \cos(\omega_s \tau) \,d\omega_s \sum_{r = -8}^8 B_r(\tau)
\ e^{i r\Omega t} \ ,
\label{eq:cycor}
\end{equation}
where
\begin{eqnarray}
{\cal P}(\omega_s) &=& \frac{48 N}{5} \ {(\omega_s L)}^4 \omega_s^{4/3}
\int_{\mathcal{M}_c} ({\sqrt{2} \mathcal{M}_c})^{10/3} P_5(\mathcal{M}_c,
\omega_s) \ d\mathcal{M}_c \ ,
\end{eqnarray}
and
\begin{equation}
B_r(\tau) = \int_{V_4} b_r(\Omega\tau)
\frac{P_4 (D,\beta,\lambda)}{D^2} \ dV_4 \ .
\label{eq:autocyc} 
\end{equation}
The functions $b_r$ entering into equation (\ref{eq:autocyc})  are
equal to
\begin{eqnarray}
b_0 &=& V_0^2 + U_0^2 +
(V_1^2 + U_1^2)\cos(  \Omega \tau)
+ U_2^2 \cos(2 \Omega \tau)
+ (V_3^2 + U_3^2)\cos(3 \Omega \tau)
\nonumber
\\ 
&&   + (V_4^2 + U_4^2)\cos(4 \Omega \tau)  +
(V_0^2 - U_0^2)\cos(4 \gamma_0) \ ,
\nonumber
\\
b_1 &=& e^{i (\Omega \tau/2 - \delta_0)}
[(U_0 U_1 + V_0 V_1)\cos(\Omega \tau/2) +
 U_1 U_2 \cos(3 \Omega \tau/2) +
\nonumber
\\ 
&& U_2 U_3 \cos(5 \Omega \tau/2)  +
(U_3 U_4 + V_3 V_4)\cos(7 \Omega \tau/2)  +
(-U_0 U_1 + V_0 V_1)\cos(\Omega \tau/2) e^{i 4 \gamma_0}]             
\ ,
\nonumber
\\
b_2 &=& e^{i (\Omega\tau - 2 \delta_0)}
[U_0 U_2 \cos(\Omega \tau) +
U_2 U_4 \cos(3 \Omega \tau) +
(V_1 V_3 + U_1 U_3)\cos(2 \Omega \tau) +
\nonumber
\\ 
&& (- U_1^2/2 + V_1^2/2 - U_0 U_2 \cos(\Omega \tau)) e^{i 4 \gamma_0}]    
\ ,
\nonumber
\\
b_3 &=& e^{i 3 (\Omega \tau/2 - \delta_0)}
[( U_0 U_3 + V_0 V_3)\cos(3 \Omega \tau/2) +
( U_1 U_4 + V_1 V_4)\cos(5\Omega \tau/2) + 
\nonumber
\\ 
&& ((-U_0 U_3 + V_0 V_3)\cos(3\Omega \tau/2)
- U_1 U_2 \cos(    \Omega \tau/2)) e^{i 4 \gamma_0}]
\ ,
\nonumber
\\
b_4 &=& e^{i 2 (\Omega \tau - 2 \delta_0)}
[(U_0 U_4 + V_0 V_4) \cos(2 \Omega \tau) +
((V_1 V_3 - U_1 U_3) \cos(\Omega\tau)  +                                   
\nonumber
\\ 
&& (V_0 V_4 - U_0 U_4) \cos(2 \Omega \tau)
- U_2^2/2) e^{i 4 \gamma_0}] 
\nonumber
\ ,
\\
b_5 &=& e^{i 5 (\Omega\tau/2 - \delta_0) +  i 4 \gamma_0}
[(V_1 V_4 - U_1 U_4)\cos(3 \Omega \tau/2) - U_2 U_3\cos(\Omega
\tau/2)]    
\ ,
\nonumber
\\
b_6 &=& e^{i (3 \Omega\tau - 6 \delta_0 + 4 \gamma_0)}
[-U_2 U_4 \cos(\Omega \tau) - U_3^2/2 + V_3^2/2]                           
\ ,
\nonumber
\\
b_7 &=& e^{i 7 (\Omega\tau/2 - \delta_0) + i 4 \gamma_0} \cos(\Omega \tau/2)
[- U_3 U_4 + V_3 V_4]                                   
\ ,
\nonumber
\\
b_8 &=& e^{i 4 (\Omega\tau - 2 \delta_0 + \gamma_0)}
[-U_4^2/2 + V_4^2/2] \ ,
\label{eqs:cofbt}
\end{eqnarray}
where $\delta_0 = \lambda - \eta_0$, $\gamma_0 = \lambda - \eta_0 -
\xi_0$, and the functions $U_i$, $V_i$ are give in equations (31-39)
of \cite{KTV04}.  It is easy to see that the autocorrelation
$B(t,\tau)$ is periodic in $t$ with period one year for a fixed
$\tau$, making it a cyclostationary random process.  Note that, if the
ecliptic longitude $\lambda$ is uniformly distributed, all the
coefficients $b_r$ given in Eq.  (\ref{eqs:cofbt}) vanish for $r > 0$,
and the random process $s(t)$ becomes stationary as the
autocorrelation $C(t',t)$ now depends on the time difference $t'-t$.

The non-stationarity of the WD-WD background was first pointed out by
Giampieri and Polnarev \cite{GP97} under the assumption of sources
distributed anisotropically, and they also obtained the Fourier
expansion of the sample variance and calculated the Fourier
coefficient for simplified WD-WD binary distributions in the
Galactic disc. What was however not realized in their work is that
this non-stationary random process is actually cyclostationary, i.e.
there exists cyclic spectra that can in principle allow us to infer
more information about the WD-WD background than one could obtain
by just estimating the zero-order spectrum.

If we now set $\tau = 0$ in Eq. \ref{eq:autocyc} we obtain the Fourier
expansion of the variance $\sigma^2(t)$ of the cyclostationary
process
\begin{equation}
\sigma^2(t) = B(t,0) =  \sum_{k = -8}^8 B_{k0} e^{i k\Omega t} \ ,
\end{equation}
where
\begin{equation}
\label{eq:sigk}
 B_{k0} = {\cal P}_o \int_{V_5} b_{k0}
\frac{P_4(D,\beta,\lambda)}{D^2} dV_4 \ ,
\end{equation}
with ${\cal P}_o = \frac{1}{2\pi}\int_0^{\infty}{\cal P}(\omega_s) \ d\omega_s$, and
\begin{eqnarray}
b_{00} &=&   U_0^2 + U_1^2 + U_2^2 + U_3^2 + U_4^2 + V_0^2 + V_1^2 +
V_3^2 + V_4^2 + 
\nonumber
\\ 
&& (V_0^2 - U_0^2) \cos(4\gamma_0)  ,\\
b_{10} &=&   e^{-i \delta_0} (U_0 U_1 + U_1 U_2 + U_2 U_3 + U_3 U_4 +
V_0 V_1 + V_3 V_4 
+ 
\nonumber
\\ 
&& (-U_0 U_1 + V_0 V_1) e^{i 4 \gamma_0}) 
\ ,
\\
b_{20} &=&   e^{-i 2 \delta_0} (U_0 U_2 + U_1 U_3 + U_2 U_4 + V_1 V_3
+
 \nonumber
\\
&& (- U_1^2/2 + V_1^2/2 - U_0 U_2) e^{i 4 \gamma_0}) 
\\
b_{30} &=&   e^{-i 3 \delta_0} (U_0 U_3 + U_1 U_4 + V_0 V_3 + V_1 V_4 + \\ \nonumber
&& (-U_0 U_3 - U_1 U_2 + V_0 V_3) e^{i 4 \gamma_0}) 
\, 
\\
b_{40} &=&   e^{-i 4 \delta_0} (U_0 U_4 + V_0 V_4 + \\ \nonumber
&& (- U_0 U_4 - U_1 U_3 + V_0 V_4 + V_1 V_3 - U_2^2/2) e^{i 4
  \gamma_0})  
\ ,
\\
b_{50} &=&  e^{i (4 \gamma_0 - 5 \delta_0)} (- U_1 U_4 - U_2 U_3 + V_1
V_4)   
\ ,
\\
b_{60} &=&  e^{i (4 \gamma_0 - 6 \delta_0)} (- U_2 U_4 - U_3^2/2 +
V_3^2/2)  
\ , 
\\
b_{70} &=&  e^{i (4 \gamma_0 - 7 \delta_0)} (- U_3 U_4 + V_3 V_4)  
\ , 
\\
b_{80} &=&  e^{i (4 \gamma_0 - 8 \delta_0)} (- U_4^2/2 + V_4^2/2) \ .
\end{eqnarray}

If we assume the function ${\cal P}(\omega_s)$ to change very little over a
frequency bin, or equivalently choose $\tau$ to be such that
$\Omega\tau \ll 1$, we can then approximate the functions $b_r$ with
the functions $b_{r0}$.  Under this approximation the cyclic spectra
of the process $s(t)$ can be shown to reduce to the following
expression
\begin{equation}
g_r(\omega_s) = \frac{1}{2}{\cal P}(\omega_s)B_{0r} \ . 
\label{eq:cycspeca}
\end{equation}
Thus under the above approximations the cyclic spectra are determined
by one function of the Fourier frequency, and by the coefficients of
the Fourier decomposition of the cyclic variance.  Note that this
simplified representation of the cyclic spectra will not be valid if
there are additional correlations between the parameters of the binary
population. For example, if the chirp masses or the frequencies of the
radiation emitted by the binaries are correlated with the positions of
the binaries themselves in the Galactic disc, then the cyclic spectra
will display a different frequency dependence from that implied by
equation (\ref{eq:cycspeca}). In general we can expect the direct
measurements of the cyclic spectra from the LISA data to allow us to
infer properties of the distribution of the parameters characterizing
the WD-WD population.  In other words, by analyzing the $17$ real and
independent cyclic spectra we should be able to derive more
information about the WD-WD binary population than we would have by
simply looking at the ordinary spectrum.

\section{Data analysis of the background signal}
\label{DATA_A}
\begin{figure}
\includegraphics[width=10cm]{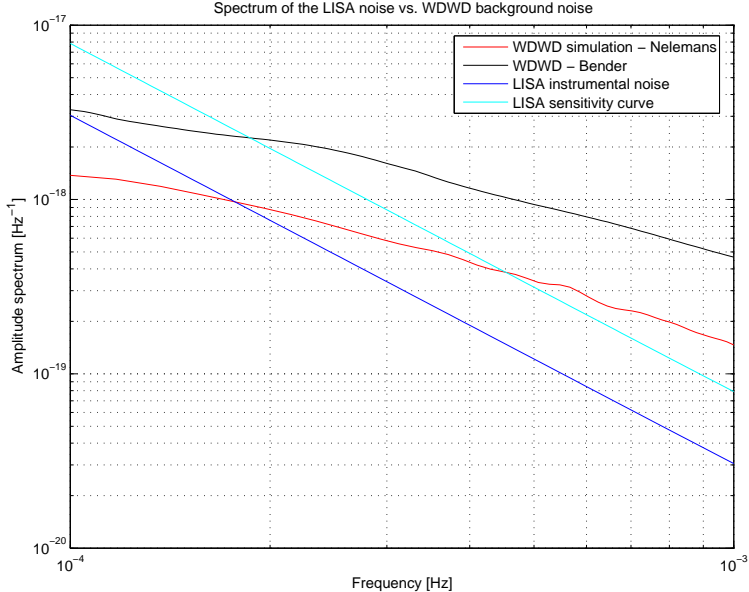}
\caption{Comparison of detached WD-WD background obtained from
  binary population synthesis simulation ( \cite{NYP01,NYP04}) with
  the WD-WD background calculated by Hils and Bender \cite{HB90}.  The
  amplitude spectral density of the LISA instrumental noise and the
  LISA sensitivity curve are drawn for comparison.  All spectral
  densities are one-sided.}
\label{fig:nel_ben_comp}
\end{figure}

We have numerically implemented the methods outlined in Section
\ref{sec:cyclo} and applied them to our simulated WD-WD background
signal.  A comparison of the results of our simulation of the detached
WD-WD background with the calculation of the background by Hils and
Bender \cite{HB90,Lweb} is shown in Figure (\ref{fig:nel_ben_comp}).
We find that the amplitude of the background from our simulation is a
factor of more than $2$ smaller than that of Hils and Bender.
The level of the WD-WD background is determined by the number of such
systems in the Galaxy. We estimate that our number WD-WD binaries should be
correct within a factor $5$ and thus the amplitude of the background should
be right within a factor of $\sqrt{5}$.
In Figure (\ref{fig:nel_ben_comp}) we have plotted the two backgrounds
against the LISA spectral density and we have also included the LISA
sensitivity curve. The latter is obtained by dividing the instrumental
noise spectral density by the detector GW transfer function averaged
over isotropically distributed and randomly polarized signals.  In the
zero-order long wavelength approximation this averaged transfer
function is equal to $\sqrt{3/20}$.

Our analysis was applied to $3$ years of LISA $X$ data consisting of a
coherent superposition of signals emitted by detached WD-WD binaries,
by semi-detached binaries (AM CVn systems), and of simulated
instrumental noise. The noise was numerically generated by using the
spectral density of the TDI $X$ observable given in \cite{ETA00}.  In
addition a $1$ mHz low-pass filter was applied to our data set in
order to focus our analyses to the frequency region in which the WD-WD
stochastic background is expected to be dominant.

\begin{figure}
\includegraphics[width=10cm]{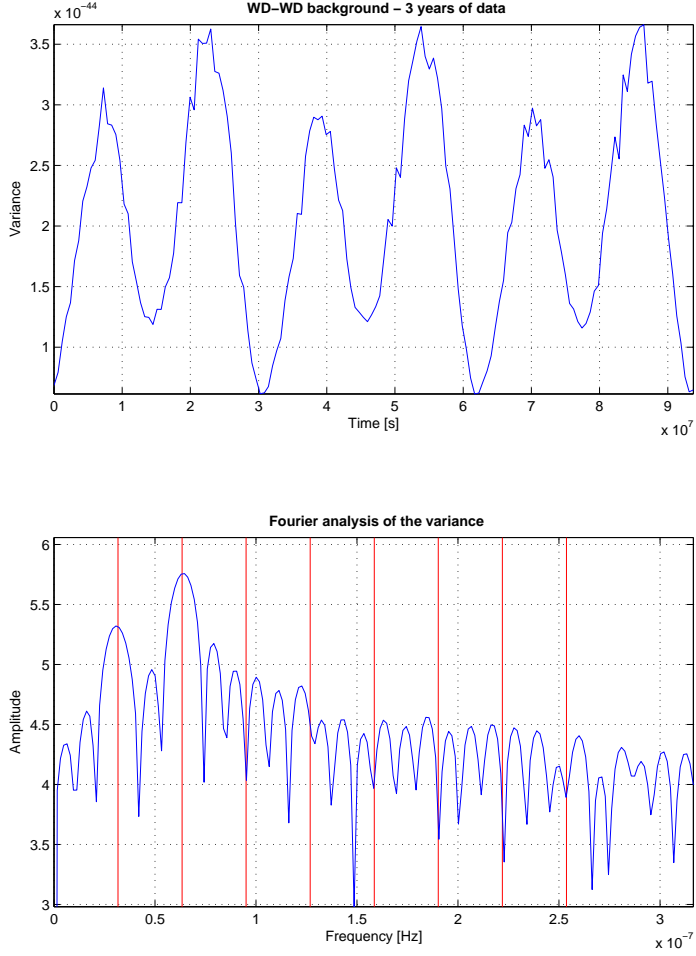}
\caption{Top panel: The sample variance of the simulated 
  WD-WD background observed by LISA. The data includes two populations
  of WD-WD binaries, detached and semi-detached, which are added to
  the LISA instrumental noise. The data is passed through a low-pass
  filter with a cut-off frequency of $1$ mHz.  Bottom panel: Fourier
  analysis of the sample variance.  Two harmonics are clearly resolved.
\label{fig:wdwd_nel_var}}
\end{figure}

\begin{figure}
\includegraphics[width=10cm]{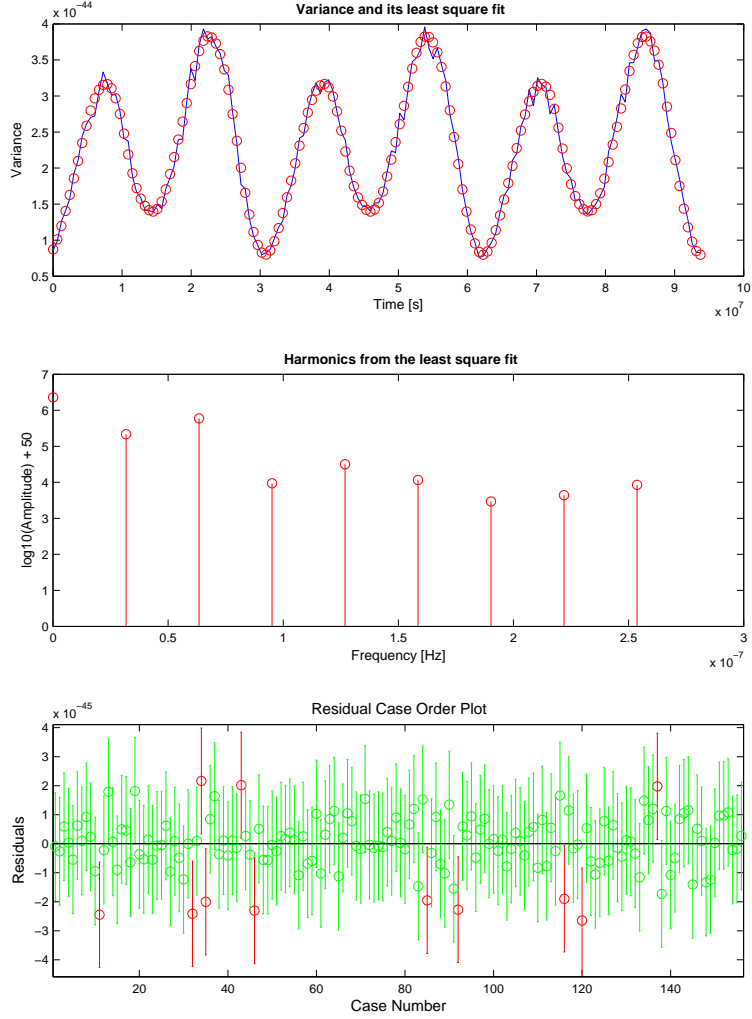}
\caption{Top panel: The sample variance of the WD-WD background data
  and the least square fit to it using $8$ harmonics (small circles).
  Middle panel: magnitude of the harmonics obtained from the least
  square fit. Bottom panel: residual error between the fit and the
  data.
\label{fig:wdwd_nel_har}}
\end{figure}

\begin{figure}
\includegraphics[width=10cm]{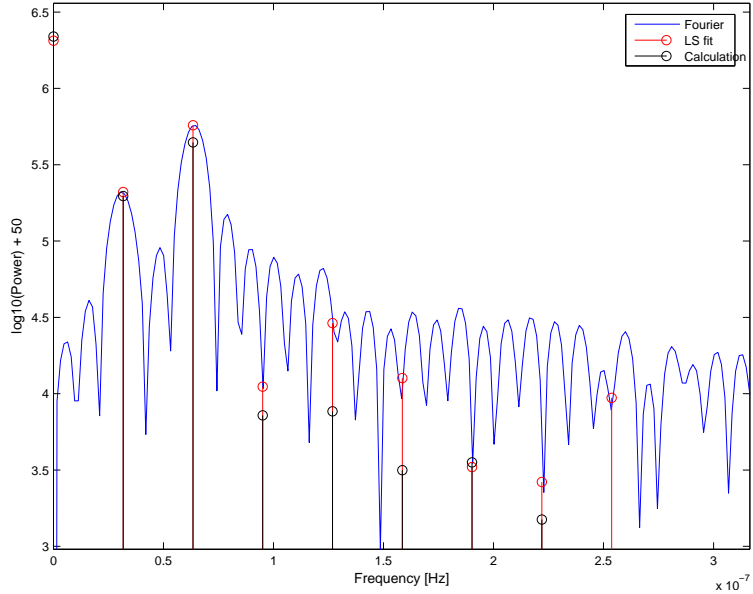}
\caption{Comparison between the estimated power in the harmonics
  obtained via (i) Fourier analysis, (ii) least square fit, and (iii)
  numerical calculation based on Eq. \ref{eq:sigk}. The blue line is
  the power spectrum of the variance of the data, red vertical lines
  are obtained from least square fit and the black vertical lines are
  from the numerical calculation.}
\label{fig:har_cyclo_comp}
\end{figure}

The results of the Fourier analysis of the sample variance of the
background signal are shown in Figures (\ref{fig:wdwd_nel_var}) and
(\ref{fig:wdwd_nel_har}). The top panel of Figure (\ref{fig:wdwd_nel_var})
shows the sample variance of the simulated data for which the
variances were estimated over a period of $1$ week; periodicity is
clearly visible.  The bottom panel instead shows the Fourier analysis
of the sample variance for which we have removed the mean from the
sample variance time series. The vertical lines correspond to
multiples of $1$ year; two harmonics can clearly be distinguished from
noise.  The other peaks of the spectrum that fall roughly half way
between the multiples of $1/{\rm year}$ frequency, are from the
rectangular window inherent to the finite time series.
In Figure (\ref{fig:wdwd_nel_har}) we present the least square fit of
$8$ harmonics to our $3$ years of simulated $X$ data. The number $8$
comes from our theoretical predictions of the number of harmonics
obtained in Section \ref{sec:cycloa} (see Eq. (\ref{eq:cycor})). We
have calculated the magnitude of the harmonics and obtained the
residuals. The results from the least square fit agree very well with
those obtained via Fourier analysis (see also Figure
(\ref{fig:har_cyclo_comp})). The magnitudes of the first and second
harmonics resolved by Fourier analysis, for instance, agree with the
corresponding least square fit estimates within a few hundredth of a
percent.

It is useful to compare the results of our numerical analysis against
the analytic calculations of Giampieri and Polnarev \cite{GP97}. Their
analytic expressions for the harmonics of the variance of a background
due to binary systems distributed in the galactic disc are given in
Eq. (42) and shown in Figure (4) of \cite{GP97}.  Our estimation
roughly matches theirs in that the 0th order harmonics is dominant and
the first two harmonics have more power than the remaining ones. Our
estimate of the power in the second harmonic, however, is larger than
that in the first one, whereas they find the opposite.  We attribute
this difference to their use of a Gaussian distribution of sources in
the Galactic disc rather than the exponential that we adopted from
\cite{NYP01}.  Comparison between these two results suggests that it
should be possible to infer the distribution of WD-WD binaries in our
Galaxy by properly analyzing the harmonics of the variance of the
galactic background measured by LISA.  How this can be done will be
the subject of a future work.

\begin{figure}
\includegraphics[width=10cm]{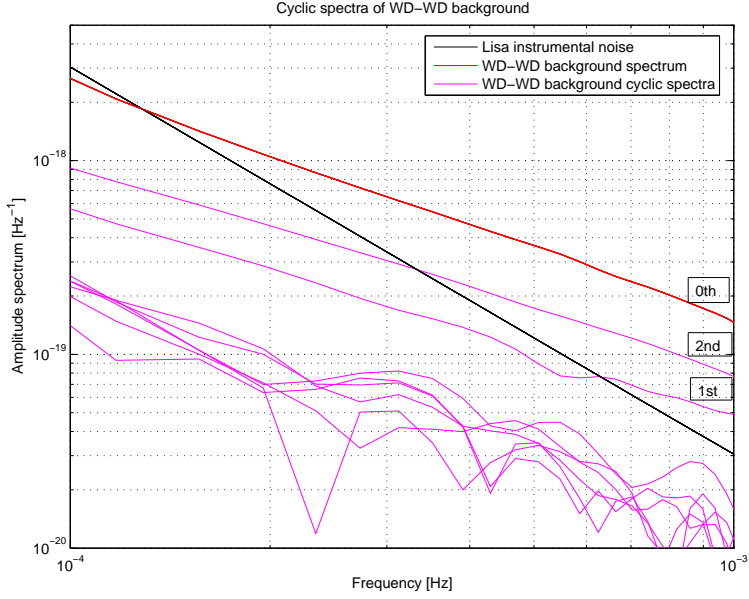}
\caption{The main (k = 0) spectrum of the 
  simulated WD-WD background signal (red), and the $8$ cyclic spectra
  (magenta) estimated from the simulated data are shown. The spectral
  density of the LISA instrumental noise (black) is shown for
  reference.
\label{fig:spec_cyclo}}
\end{figure}

\begin{figure}
\includegraphics[width=10cm]{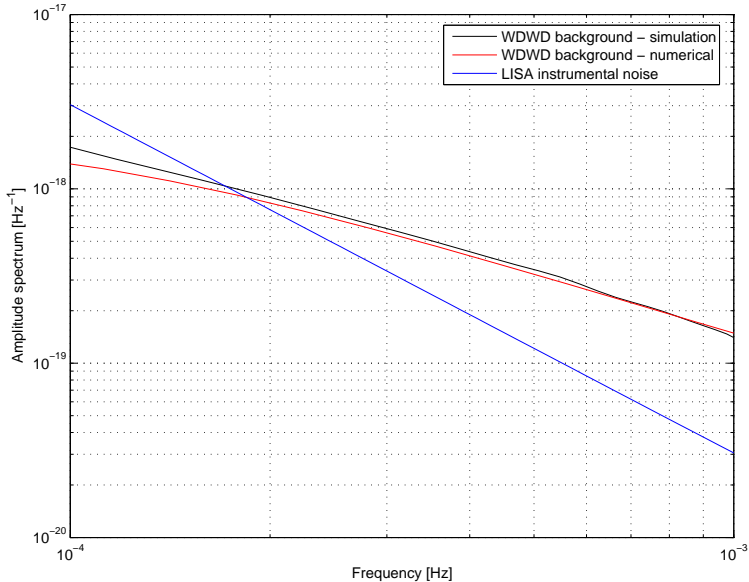}
\caption{Estimated main ($k = 0$) spectrum of the WD-WD background
  (red) against the calculated spectrum (black). The LISA spectral
  density curve (blue) is shown for comparison. The $0$th order
  spectrum contains the LISA instrumental and hence it differs from the
  spectrum given in Figure (\ref{fig:nel_ben_comp}).
\label{fig:spec_comp}}
\end{figure}

\begin{figure}
\includegraphics[width=10cm]{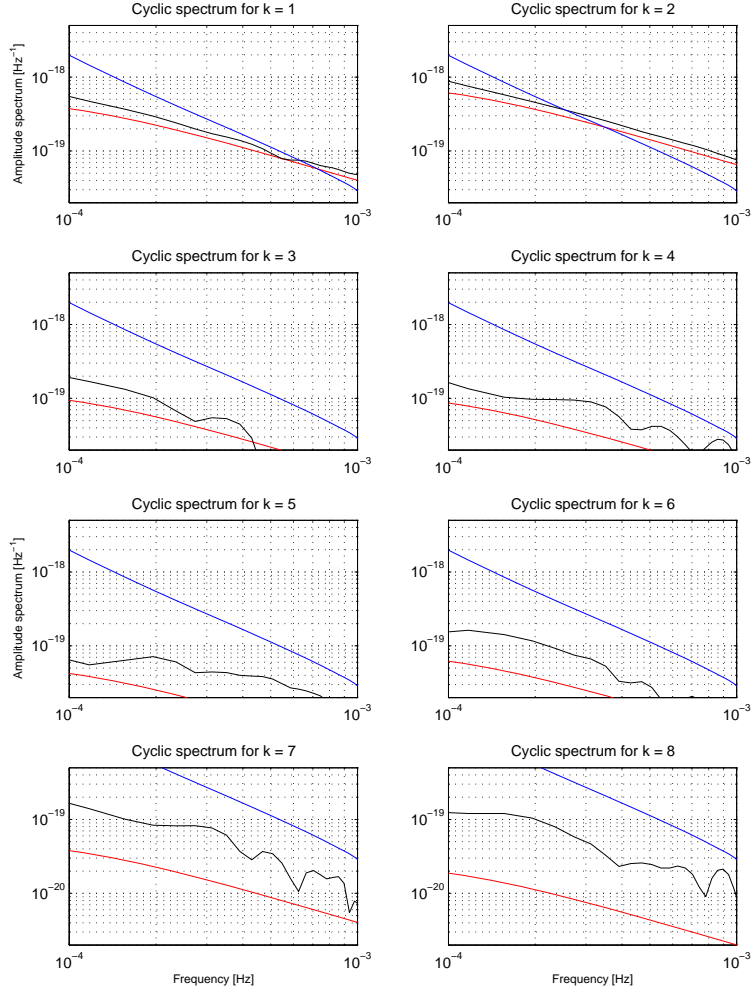}
\caption{Estimated cyclic spectra (black) against the calculated spectra (red).
  The LISA spectral density curve (blue) is shown for comparison.
\label{fig:spec_cyclo_comp}}
\end{figure}

In order to validate our simulation and data analysis method we have
compared the results of our estimation of the power in the harmonics
of the variance against the explicit analytic calculation. To estimate
the powers we have used Eq. (\ref{eq:sigk}) and we have evaluated the
integrals by numerical and Monte Carlo methods. In the numerical
calculation of the harmonics we have limited our analysis to the
population of detached WD-WD binaries. Thus in order to make the
comparison meaningful we have performed Fourier analysis and least
square fit of the time series consisting only of simulated detached
WD-WD binaries (without semi-detached ones and LISA instrumental
noise). The results of the comparison are given in Figure
(\ref{fig:har_cyclo_comp}). We see that for the 0th order harmonic and
the first two harmonics the agreement is very good.  For higher order
harmonics there are large discrepancies between the numerical
calculation and estimation by the least square fit, while by using the
Fourier transform method, we cannot even resolve higher harmonics in
our $3$-year data set. We conclude that only the two first harmonics
can be extracted reliably from the data. We also observe a very good
agreement between the Fourier and the least square method.
As a next step in our analysis, we have estimated the cyclic spectra
of the simulated WD-WD background signal. In Figure
(\ref{fig:spec_cyclo}) we have shown the cyclic spectra estimated from
the data.  We have also plotted the spectrum of the LISA instrumental
noise and the main spectrum ($k = 0$) estimated from the simulation.
We find that the main spectrum and two cyclic spectra for $k = 1$ and
$k = 2$ have the largest magnitude and, over some frequency range,
they lie above the LISA instrumental noise. The remaining spectra are
an order of magnitude smaller and are very noisy. We also see that all
the cyclic spectra have roughly the same slope. This is predicted by
our analytic calculations in Section \ref{sec:cycloa} and it follows
from the assumed independence between the location of the binaries in
the Galaxy ($D, \lambda, \beta$) and their frequencies and chirp
masses ($\omega_s, {\cal M}_c$). We also find the magnitude of the 2nd
cyclic spectrum to be higher than the first, similarly to what we had
for the harmonics of the variance. Note that we estimated the spectra from 
the time series consisting of the WD-WD background added to 
the LISA instrumental noise.
Like the analysis we did for the variance, we have also compared the
estimates of the cyclic spectra from our simulation against those
obtained via numerical calculation of the equations derived in Section
\ref{sec:cycloa}. The corresponding results are presented in Figures
(\ref{fig:spec_comp}) and (\ref{fig:spec_cyclo_comp}), where it is shown
that the agreement between the two is quite good.


Our analysis has shown that the LISA data will allow us to compute
$17$ independent cyclic spectra (the $8$ complex cyclic spectra
$g_r(f) , r=1, 2, ...8$ and the real spectrum $g_0 (f)$) of the WD-WD
galactic background, $5$ of which can be expected to be measured
reliably.  We have also shown that by performing generalized spectral
analysis of the LISA data we will be able to derive more information
about the WD-WD binary population (properties of the distribution of
its parameters) than we would have by only looking at the ordinary
$g_0 (f)$ spectrum.

\section*{Acknowledgments}
The supercomputers used in this investigation were provided by funding
from the Jet Propulsion Laboratory Institutional Computing and
Information Services, and the NASA Directorates of Aeronautics
Research, Science, Exploration Systems, and Space Operations.  A.K.\ 
acknowledges support from the National Research Council under the
Resident Research Associateship program at the Jet Propulsion
Laboratory. This work was supported in part by the Polish Science
Committee Grant No. KBN 1 P03B 029 27.  This research was performed at
the Jet Propulsion Laboratory, California Institute of Technology,
under contract with the National Aeronautics and Space Administration.

\end{document}